\documentclass[a4paper]{article}
\usepackage[margin=4cm]{geometry}

\usepackage{amssymb,amsmath,amsfonts}
\usepackage{amsthm} 
\usepackage{graphicx} 
\usepackage{lmodern}
\usepackage{todonotes}
\usepackage{url}
\usepackage{pdfsync}
\usepackage{paralist}
\usepackage{bbm}
\usepackage{algorithm}
\usepackage{algorithmic}

\usepackage{algorithm}
\usepackage{algorithmic}

\newcommand{\p}{{\mathbf{p}}}

\newcommand{\Ccal}{{\mathcal C}}

\newcommand{\Xcal}{{\mathcal X}}
\newcommand{\Acal}{{\mathcal A}}
\newcommand{\Pcal}{{\mathcal P}}
\newcommand{\Mcal}{{\mathcal M}}

\newcommand{\ranktt}{{\text{rank}_{\text{TT}}}}
\newcommand{\vc}{{\text{vec}}}
\makeatletter
\def\BState{\State\hskip-\ALG@thistlm}
\makeatother

\newcommand{\size}{{\rm size}}

\newcommand{\norm}[1]{\lVert #1 \rVert}

\newcommand{\defby}{\mathrel{\mathop:}=}
\newcommand{\bydef}{=\mathrel{\mathop:}}

\newtheorem{proposition}{Proposition}[section]

\newtheorem{remark}[proposition]{Remark}

\begin{document}

\title{Low-rank tensor approximation for Chebyshev interpolation in parametric option pricing\footnote{The authors would like to thank Jonas Ballani for helpful discussions on this work.}
}

\author{Kathrin Glau\footnote{Queen Mary University of London, Mile End Road, E1 4NS London, United Kingdom, {\tt k.glau@qmul.ac.uk}} \and Daniel Kressner\footnote{\'Ecole Polytechnique F\'ed\'erale de Lausanne, Station 8, 1015 Lausanne, Switzerland ({\tt daniel.kressner@epfl.ch, http://anchp.epfl.ch)}} \and Francesco Statti\footnote{\'Ecole Polytechnique F\'ed\'erale de Lausanne, Station 8, 1015 Lausanne, Switzerland ({\tt francesco.statti@epfl.ch, http://people.epfl.ch/francesco.statti).} Research supported through the European Research Council under the European Union\textsc{\char13}s Seventh Framework Programme (FP/2007-2013) / ERC Grant Agreement n. 307465-POLYTE.}}

\date{February 11, 2019}

\maketitle

\begin{abstract}
Treating high dimensionality is one of the main challenges in the development of computational methods for solving problems arising in finance, where tasks such as pricing, calibration, and risk assessment need to be performed accurately and in real-time.
Among the growing literature addressing this problem, Gass et al.~\cite{gass2016chebyshev} propose a complexity reduction technique for parametric option pricing based on Chebyshev interpolation. As the number of parameters increases, however, this method is affected by the curse of dimensionality. In this article, we extend this approach to treat high-dimensional problems: Additionally exploiting low-rank structures allows us to consider parameter spaces of high dimensions. The core of our method is to express the tensorized interpolation in tensor train (TT) format and to develop an efficient way, based on tensor completion, to approximate the interpolation coefficients. 
We apply the new method to two model problems: American option pricing in the Heston model and European basket option pricing in the multi-dimensional Black-Scholes model. In these examples we treat parameter spaces of dimensions up to 25. The numerical results confirm the low-rank structure of these problems and the effectiveness of our method compared to advanced techniques.
\end{abstract}


\section{Introduction}\label{section1}
Financial problems are, by their nature, multi- and high-dimensional, because a large number of risk factors contribute to the prices of each financial asset. Moreover, the banking, insurance and hedge fund industry draws on investments in large portfolios. 
The interdependencies of both the risk factors and the assets make basic computational tasks such as model calibration, pricing, and hedging as well as more global tasks such as uncertainty quantification, risk assessment and capital reserve calculation computationally extremely challenging, see for instance \cite{BarreraCrepeyDialloFortGobetStazhynski2019}.

Automatic and high-speed trading challenge the computational methods in that the results need to be available fast and with minimal storage requirement. Moreover, we observe  rising regulatory requirements. On the one hand, more realistic modeling demands more prudent considerations, which leads to rising computational complexity. On the other hand, the availability of requested performance characteristics is expected to be delivered within shorter periods of time. This poses a high challenge for traditional approaches, which typically suffer from low convergence rates in higher dimensions, see for instance \cite{CapriottiJiangMacrina2017, dempster2018high}.

For the reasons explained above, the development of efficient computational methods for high-dimensional problems in finance is an utmost active field of research in both academia and industry. For example, further developments of the Monte Carlo method have been very successfully applied to financial problems; we refer to \cite{LEcuyer2009, GilesXia2017} for the quasi Monte Carlo method and to \cite{Giles2015} for the multilevel Monte Carlo method. Besides stochastic integration, deterministic numerical integration has been exploited using sparse grid techniques, see~\cite{griebel2010dimension, holtz2011sparse, BayerSiebenmorgenTempone2017}.
Also PDE methods have been extended to multivariate problems in finance. For instance using operator splitting methods as in~\cite{HoutToivanen2016}, principal component analysis and
expansions as in~\cite{ReisingerWissmann2016}, and wavelet compression techniques proposed in~\cite{MatacheNitscheSchwab2005,HilberReichSchwabWinter2009,HilberReichmannSchwabWinter2013}. 
  
Exploiting the particular structure of a problem, \emph{complexity reduction techniques} exhibit great potential to save run-time and storage capacity while maintaining the required accuracy. In numerical analysis and a large variety of applications, for example in engineering and medicine, complexity reduction techniques have been developed and implemented with great success. For instance the field of reduced basis methods to efficiently solve parametric partial differential equations (PDEs) has experienced a tremendous development over the last decade, see, e.g.,~\cite{Hesthaven2015, Patera2006, QuarteroniManzoniNegri2016} and the references therein. Pioneered by  \cite{SachsSchu2008, ContLantosPironneau2011} the potential of reduced basis methods is also increasingly exploited for problems in finance; see~\cite{BurkovskaHaasdonkSalomonWohlmuth2015,MayerhoferUrban2017,BurkovskaGlauMahlstedtWohlmuth2017} for examples.
 These methods can be viewed as high-dimensional interpolation methods that are trained in an offline step to solve a specific class of parametric PDEs. In this article we explore direct interpolation of multivariate functions as a \emph{unified approach} to complexity reduction for finance.

Our starting point is the tensorized Chebyshev interpolation of conditional expectations in the parameter and state space, as introduced in~\cite{gass2016chebyshev}. Having observed for a large set of applications that these functions are highly regular, admitting sensitivities of high order or even being analytic, and that the domain of interest can be restricted to a hyperrectangular, Chebyshev interpolation is a promising choice: Its convergence is subexponential for multivariate analytic functions, its implementation is numerically stable, and the coefficients are simply given by a linear transformation of the function values at the nodal points. In this article we exploit this favorable structure further for high dimensionality. In passing, we point out that, while we choose Chebyshev interpolation for the reasons listed in this paragraph, the technique presented in this paper extends to other tensorized interpolation techniques. Also, our approach is applicable beyond option pricing and finance.

The basis of our approach is the following. In an \emph{offline phase}, the price as function of parameters $\p\in[-1,1]^d$, $\p\mapsto \mathsf{Price}^\p$ is evaluated at selected parameter samples $\p$ to prepare an approximation by tensorized Chebyshev polynomials $ T_{j_1, \dots, j_d}$ with pre-computed Fourier coefficients $c_{j_1, \dots, j_d}$, as follows,
	\begin{equation}\label{eq-priceapprox}
		\mathsf{Price}^\p \approx  \sum_{j_1=0}^{n_1} \dots \sum_{j_d=0}^{n_d} c_{j_1, \dots, j_d} T_{j_1, \dots, j_d}(\p).
	\end{equation}
To evaluate the function in the \emph{online phase}, only the multivariate polynomials on the right-hand side need to be evaluated. However, implementing \eqref{eq-priceapprox} in a straightforward manner exposes the method to the curse of dimensionality in both the offline and the online phase: In the offline phase, the prices need to be evaluated on a tensorized grid of Chebyshev nodes, amounting to $O(n^d)$ parameter samples when $n$ nodes are required for each parameter. This is computationally costly, especially if the underlying pricing method is already computationally demanding. In the online phase alike, $O(n^d)$ operations are needed for evaluating the approximating multivariate polynomial. Even for a number as low as $n=3$, corresponding to quadratic polynomials, a problem with $d=20$ parameters becomes infeasible.

One approach to breaking the curse of dimensionality that has already proven effective in a number of areas is to exploit low-rank structures of high-dimensional tensors; see \cite{grasedyck2013,hackbusch2012tensor,Khoromskij2018} and the references therein. These techniques reduce, sometimes dramatically, memory requirements and the cost of operating with tensors. In the context of parametric PDEs, low-rank tensor structures have been successfully exploited in, e.g.,~\cite{Bachmayr2017,Ballani2015,Khoromskij2011c,kressner2011low-rank,steinlechner2016riemannian}. As option prices are characterized as solutions of parabolic PDEs, this gives hope that low-rank structures can be exploited in finance as well. The following questions arise:

\textit{Can we detect low-rank structures for the problem of form \eqref{eq-priceapprox}?} Existing theoretical studies only provide partial answers to this question, either not reflecting the observed effectiveness of low-rank techniques or being limited to rather specific function classes; see~\cite{Dahmen2016,hackbusch2012tensor,SchneiderUschmajew2014} for examples. We therefore approach the question from an experimental perspective and analyze examples of different nature and different dimensionality in Section~\ref{section3}. The results clearly indicate an approximate low-rank structure of the tensor $\mathcal{P}$ containing the prices evaluated at the nodes of the tensorized Chebyshev grid.
In the specific case of the interpolation of American option prices in the Heston model in five parameters we can explicitly compare the full tensor $\mathcal{P}$ with the one  resulting from low-rank approximation. We perform this comparison in Section~\ref{sectionHeston}, which confirms the low-rank structure of $\mathcal{P}$. In Section~\ref{numerical exp BS} we consider prices of basket options in the Black-Scholes model with up to $25$ underlyings and interpolate in the initial values of the underlyings. Although the resulting full tensor $\mathcal{P}$ is too large to be explicitly computed and compared with, we provide a structural analysis in Section~\ref{analysisP} that explains why $\mathcal{P}$ is expected to exhibit low-rank structure.

\textit{How can we exploit low-rank structures for the problem of form \eqref{eq-priceapprox}?} Expressing the problem in a tensor format reveals that exploiting the tensor structure itself (even without low-rank structure) leads to a considerable efficiency gain in both the offline and the online phase. Next, we explore existing low-rank tensor techniques. In order to efficiently exploit these techniques for problem \eqref{eq-priceapprox}, we need to introduce several new components resulting in the new method.
 We detail these steps below.

In order to construct the interpolation coefficients $c_{j_1, \dots, j_d}$ in the offline phase, it is first required to compute or approximate all values of the tensor $\mathcal{P}$, containing the prices in the tensorized Chebyshev grid. Evaluating $\mathcal{P}$ explicitly is too costly for larger $d$, especially when the underlying pricing procedure is computationally expensive.
Instead we only compute part of the entries of $\mathcal{P}$ and then need to deal with an incomplete tensor.
This leads us to the following first step: 
\begin{itemize}
\item[1.] We start by computing the prices for a small portion of the Chebyshev grid points only. 
Then, we adapt a \emph{completion algorithm} (in Section \ref{sec-completion}) which allows us to approximate the tensor of prices for the complete Chebyshev grid by fitting tensors of pre-specified low rank to the provided data points. As it is not reasonable to assume a priori knowledge of low-rank structure, 
the completion procedure needs to be combined with an \emph{adaptive rank and sampling strategy}. Specifically, we
 repeat the process of adding new samples and increasing the pre-specified rank until an adequate stopping criterion is fulfilled. This completion algorithm is designed to work with tensors built and stored in tensor train (TT) format.
 \end{itemize}
With the low-rank approximation of the tensor $\Pcal$ in TT format at hand, we can then approximate efficiently the Fourier coefficients $c_{j_1, \dots, j_d}$. This is the last step of the offline phase:
\begin{itemize}
\item[2.] 
The computation of the tensor $\mathcal{C}$, containing the Fourier coefficients $c_{j_1, \dots, j_d}$, is computed by a sequence of $d$ tensor-matrix multiplications. The particular structure of the involved matrices facilitates the use of the fast Fourier transform, leading to a complexity of  $O(dnr^2\log(n))$, where $r$ is determined by the ranks of $\mathcal{P}$. This step is explained in Section \ref{computationC}.
\end{itemize}
Suppose now that, in the online phase, we want to compute the interpolated price \eqref{eq-priceapprox} for a new set of parameter samples. Given the tensor $\mathcal{C}$ in TT format, the evaluation of \eqref{eq-priceapprox} for a price $\p$ is performed efficiently as follows:
\begin{itemize}
\item[3.]
First, each of the Chebyshev polynomials involved in the tensorized Chebyshev basis is evaluated in $\p$. It turns out that~\eqref{eq-priceapprox} can be viewed as inner product between $\mathcal{C}$ and a rank-one tensor. Thanks to the TT format, the complexity of computing this inner product is $O(dnr^2)$;  see Section \ref{sec-ttformat}. As long as $r$ is reasonably small, this compares favorably with the $O(n^d)$ operations needed by the standard approach. 
\end{itemize}

In Section~\ref{section3}, we test the performance of the new method for two different option pricing problems, the interpolation of
\begin{itemize}
\item[--] American option prices in the Heston model in $d=5$ parameters, and of 
\item[--] prices of basket options in the Black-Scholes model in up to $d=25$ underlyings.
\end{itemize}
At comparable accuracy, the interpolation in American option prices reveals a promising gain in efficiency when compared to an ADI-based PDE solver. The efficiency gain for the basket option prices is shown in comparison to a Monte Carlo simulation with variance reduction.

\section{TT format and tensor completion for Chebyshev interpolation}\label{section2}

This section describes the methodology proposed in this work. We start with recalling the tensorized Chebyshev interpolation method from~\cite{gass2016chebyshev}. After introducing the TT format~\cite{oseledets2011}, we present and extend the tensor completion approach from~\cite{steinlechner2016riemannian}. Finally, we explain how to combine these algorithms in order to efficiently price parametric options for a large number of parameters.

\subsection{Chebyshev interpolation for parametric option pricing}\label{chebyshevPOP}
We consider an option price that depends on a vector of $d$ parameters $\mathbf{p}$ contained in $[-1,1]^d$; general hyperrectangular parameter domains can be addressed by a suitable affine transformation. The basic idea developed in~\cite{gass2016chebyshev} consists of using tensorized Chebyshev interpolation in the parameters (model and payoff parameters) to increase the efficiency of computing option prices, while maintaining satisfactory accuracy. Writing $\mathsf{Price}^\p$ for the price evaluated in $\p$, the Chebyshev interpolation of order $\overline{\mathbf{n}}:=(n_1, \dots, n_d)$ with $n_i \in \mathbb{N}_0$ is given by
\begin{equation}\label{intprice}
I_{\overline{\mathbf{n}}}(\mathsf{Price}^{(\cdot)})(\p)= \sum_{j_1=0}^{n_1} \dots \sum_{j_d=0}^{n_d} c_{j_1, \dots, j_d} T_{j_1, \dots, j_d}(\p).
\end{equation}
The basis functions $T_{j_1, \dots, j_d}$ are constructed from Chebyshev polynomials by
\begin{equation}\label{cheb-basis}
T_{j_1, \dots, j_d}(\p)= \prod_{i=1}^d T_{j_i} (p_i),\quad T_{j_i} (p_i)=\cos(j_i \arccos(p_i)),
\end{equation}
and the coefficients $c_{j_1, \dots, j_d}$ are defined as
\begin{equation}\label{coeff}
c_{j_1, \dots, j_d}=\Big( \prod_{i=1}^{d} \frac{2 ^{\mathbbm{1}_{n_i>j_i >0}}}{n_i}\Big) \sideset{}{''}\sum_{k_1=0}^{n_1} \dots \sideset{}{''}\sum_{k_d=0}^{n_d} \Pcal (k_1,\dots, k_d) \prod_{i=1}^d \cos \big(j_i \pi \frac{k_i}{n_i}   \big),
\end{equation}
where $\sideset{}{''}\sum $ indicates that the first and the last summand are halved. The tensor $\Pcal$ contains the prices on the tensorized Chebyshev grid:
\[
\Pcal (k_1,\dots, k_d) = \mathsf{Price}^{q_{k_1,\dots, k_d}},
\]
where $q_{k_1,\dots, k_d}:=(q_{k_1}, \dots,q_{k_d} )$ is defined via Chebyshev nodes $q_{k_i}:= \cos(\pi \frac{k_i}{n_i})$ for $k_i=0 ,\dots, n_i$ and $i=1 ,\dots, d$.  A convergence analysis of the tensorized Chebyshev interpolation in the setting of option pricing is given in~\cite{gass2016chebyshev}.

The tensor $\mathcal{P}$ in equation \eqref{coeff} is of order $d$ and size $(n_1+1) \times \dots \times (n_d+1)$. The interpolation procedure first requires to compute each entry of this tensor with the reference method. This becomes expensive when the interpolation order and the dimension $d$ increase. We will use tensor completion to lower this cost.
\begin{remark} [Choice of interpolation order]
In our numerical experiments, the interpolation order $\overline{\mathbf{n}}$ is chosen a priori for simplicity. However, this choice can be made adaptively as explained in~\cite{hashemi2016chebfun} for the case $d=3$ (the extension to general $d$ is straightforward). \end{remark}

\subsection{TT format}\label{sec-ttformat}

\begin{figure}[t]
\centering
\includegraphics[width=0.7\textwidth]{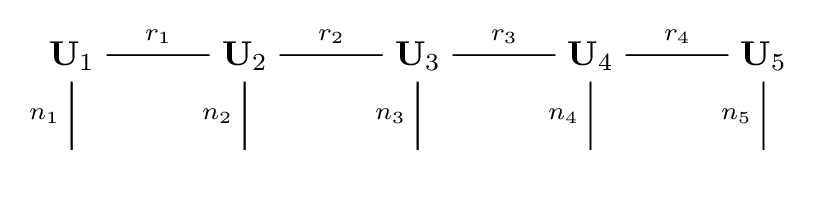}
\caption{Tensor network diagram of TT decomposition for a tensor of order $d=5$. \label{ttrepresentation}}
\end{figure}

For recalling the TT format introduced in \cite{oseledets2011}, we consider a general tensor $\mathcal{X} \in \mathbb{R}^{n_1 \times n_2 \times \cdots \times n_d}$ of order $d$. For each $\mu = 1,\ldots,d-1$, the entries of $\mathcal{X}$ can be rearranged into a matrix
\begin{equation*}
X^{<\mu>} \in \mathbb{R}^{(n_1n_2 \cdots n_{\mu}) \times (n_{\mu + 1}\cdots n_d)},
\end{equation*}
which is called the \textsl{$\mu$th unfolding} of $\mathcal{X}$. For this purpose, the first $\mu$ indices of $\mathcal{X}$ are merged into the row index and the last $n-\mu$ indices into a column index; see~\cite{oseledets2011} for a formal definition. The \textsl{TT ranks} of $\mathcal{X}$ form an integer tuple \begin{equation}\label{ttrankdef}
\ranktt (\mathcal{X})= (r_0,r_1,\cdots,r_d):=(1, \text{rank}(X^{<1>}), \cdots, \text{rank}(X^{<d -1>}), 1).
\end{equation}
Every entry $\mathcal{X}(i_1, i_2, \cdots, i_d)$ can be expressed as a product of $d$ matrices
\begin{equation*}
\mathcal{X}(i_1, i_2, \cdots, i_d)=U_1(i_1)U_2(i_2) \cdots U_d(i_d),
\end{equation*}
with $U_{\mu}(i_{\mu})$ a matrix of size $r_{\mu-1} \times r_{\mu}$. For each $\mu=1,\cdots,d$, one can then collect the $n_{\mu}$ matrices $U_{\mu}(i_{\mu})$, $i_{\mu}=1,2,\cdots, n_{\mu}$ into a third order tensor $\mathbf{U}_{\mu}$ of size $r_{\mu-1} \times n_{\mu} \times r_{\mu}$. These tensors are called \textsl{TT cores} and, by construction, we have
\begin{equation}\label{TTformat2}
\mathcal{X}(i_1, i_2, \cdots, i_d)=\sum_{k_1=1}^{r_1} \cdots \sum_{k_{d-1}=1}^{r_{d-1}}  \mathbf{U}_{1}(1,i_1,k_1) \mathbf{U}_{2}(k_1,i_2,k_2) \cdots
\mathbf{U}_{d}(k_{d-1},i_d,1).
\end{equation}
Figure \ref{ttrepresentation} illustrates this so-called \textsl{TT decomposition} by a tensor network diagram~\cite{orus2014practical}.
Provided that the TT ranks remain moderate, a significant memory reduction is obtained by storing instead of $\mathcal{X}$ the TT cores: from $O(n^d)$ to $O(dnr^2)$, where $r = \max\{r_0,\ldots,r_d\}$ and $n = \max\{n_1,\ldots,n_d\}$.

\begin{figure}[ht]
\centering
\includegraphics[width=0.8\textwidth]{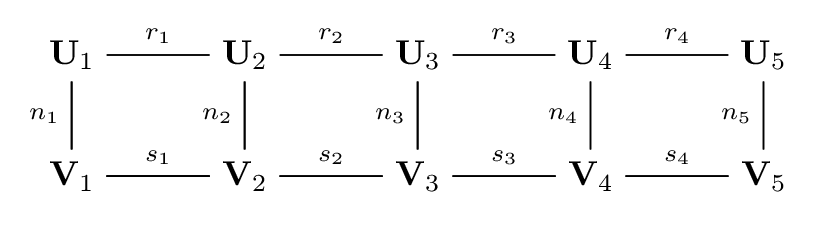}
\caption{Inner product of two tensors of order $d=5$ in TT decomposition. \label{ttinner}}
\end{figure}

Some operations can be effected quite cheaply in the TT format for tensors of low TT ranks. Let us first consider the inner product of
two tensors $\mathcal{X}, \mathcal{Y} \in \mathbb{R}^{n_1 \times \cdots \times n_d}$ defined as
\begin{equation}\label{innerproduct}
\langle \mathcal{X}, \mathcal{Y} \rangle = \langle \vc (\mathcal{X}) , \vc(\mathcal{Y}) \rangle = \sum_{i_1 = 1}^{n_1} \cdots \sum_{i_d = 1}^{n_d} \mathcal{X}(i_1,\ldots,i_d)\mathcal{Y}(i_1,\ldots,i_d),
\end{equation} 
where $\vc(\cdot)$ stacks the entries of a tensor into a long vector. The corresponding tensor network diagram when $\mathcal X$ and $\mathcal Y$ are both in TT decomposition is shown in Figure \ref{ttinner}. It can be seen that the summations in~\eqref{innerproduct} become contractions between the TT cores of $\mathcal X$ and $\mathcal Y$. By carrying out these contractions of cores from the left to right, the cost of evaluating the inner product reduces from $O(n^d)$ to $O(dnr^3)$, where $r$ denotes the maximum of all involved TT ranks. 

The \textsl{mode-$\mu$ matrix multiplication} between 
a tensor $\mathcal{X}\in \mathbb{R}^{n_1 \times \cdots \times n_d}$ and a matrix $M \in \mathbb{R}^{m \times n_\mu}$  results in a tensor $\mathcal{Z} \in \mathbb{R}^{n_1 \times \cdots n_{\mu-1} \times m \times n_{\mu+1} \cdots \times n_d}$ defined by 
\begin{equation*}
\mathcal{Z}(i_1, \cdots ,i_{\mu-1},j, i_{\mu+1} \cdots, i_d)=\sum_{i_\mu=1}^{n_\mu}
\mathcal{X}(i_1, \cdots, i_d) M(j, i_k), \quad j = 1,\ldots, m.
\end{equation*}
We will denote this operation by $\mathcal{Z}=\mathcal{X} \times_k M$. If $\mathcal X$ is in TT decomposition~\eqref{TTformat2} then it is straightforward to obtain a TT decomposition for $\mathcal Z$, by performing a mode-$2$ matrix multiplication of $\mathbf U_\mu$ with $M$. Letting $c_M$ denote the cost of multiplying $M$ with a vector, this requires $O(c_M n r)$ operations instead of the $O(c_M n^{d-1})$ operations needed when $\mathcal X$ is a general tensor.

\subsection{Completion algorithm}\label{sec-completion}

The goal of completion algorithms is to reconstruct a given data set from a small fraction of its entries. As this is clearly an ill-posed task, one needs to  additionally impose some regularization, such as smoothness conditions. In this work, we impose low TT ranks on the tensor $\Pcal$ containing the prices and reconstruct $\Pcal$ using the completion algorithm proposed in~\cite{steinlechner2016riemannian}.

In the following, we briefly summarize the approach from~\cite{steinlechner2016riemannian}. Let $\Acal \in \mathbb{R}^{n_1 \times \cdots \times n_d}$ denote the original data tensor for which only the entries in a (small) \textsl{training set} $\Omega \subset \{1,n_1\} \times \cdots \times \{1, n_d\}$ are known. When aiming at fitting a tensor of fixed (low) TT ranks $\mathbf{r} = (r_0,\ldots,r_d)$ to this data,  completion takes the form of the constrained optimization problem
\begin{align} \label{completionproblem}
\begin{split}
\min_{\Xcal} \quad & ||P_{\Omega} \Xcal-P_{\Omega} \Acal||^2\\
\text{subject to} \quad & \Xcal \in \Mcal_\mathbf{r} := \{\Xcal \in \mathbb{R}^{n_1 \times \cdots \times n_d} \enskip | \enskip \ranktt= \mathbf{r}\},
\end{split}
\end{align}
where $P_{\Omega} \Xcal$ denotes the orthogonal projection onto $\Omega$ and $\|\cdot\|$ is the norm induced by the inner product~\eqref{innerproduct}. It is known that $\Mcal_\mathbf{r}$ is a smooth embedded submanifold, which enables one to apply Riemannian optimization techniques to~\eqref{completionproblem}. Specifically, in~\cite{steinlechner2016riemannian} it is proposed to employ a Riemannian conjugate gradient (CG) method (see Algorithm 1 in~\cite{steinlechner2016riemannian}). This method produces iterates that stay on the manifold and, in turn, can be stored and manipulated efficiently in the TT format. One iteration requires $O(dnr^3+d|\Omega|r^2)$ operations, where $|\Omega|$ denotes the cardinality of $\Omega$. 

Our stopping criterion of Riemannian CG is designed to attain a level of accuracy warranted by the data and the chosen TT ranks. Following~\cite{steinlechner2016riemannian}, we choose a \textsl{test set} ${\Omega_C}$ of, say, $100$ additional parameter samples not in the training set $\Omega$. Letting $\Xcal_k$ denote the $k$th iterate of Riemannian CG algorithm, we measure the errors on the training and the test set:
\begin{equation*}
\epsilon_{\Omega} (\Xcal_k)\defby \frac{\norm{P_{\Omega}  \mathcal{A}-P_{\Omega} \Xcal_k}}{\norm{P_{\Omega} \mathcal{A}}}, \quad
\epsilon_{{\Omega_C}} (\Xcal_k)\defby \frac{\norm{P_{{\Omega_C}}  \mathcal{A}-P_{{\Omega_C}} \Xcal_k}}{\norm{P_{{\Omega_C}} \mathcal{A}}}.
\end{equation*}
The algorithm is stopped once these errors stagnate, that is,
\begin{equation}\label{stoppingCG}
\frac{|\epsilon_{\Omega}(\Xcal_k)-\epsilon_{\Omega}(\Xcal_{k+1})|}{|\epsilon_{\Omega}(\Xcal_k)|} < \delta \quad \text{and} \quad 
\frac{|\epsilon_{{\Omega_C}}(\Xcal_k)-\epsilon_{{\Omega_C}}(\Xcal_{k+1})|}{|\epsilon_{{\Omega_C}}(\Xcal_k)|} < \delta,
\end{equation}
holds for some small $\delta > 0$. 
\subsubsection{Adaptive rank and adaptive sampling strategy}
 
To set up the optimization problem~\eqref{completionproblem}, two issues remain to be discussed: The choice of the TT ranks $\mathbf{r}$ and a suitable training set $\Omega$. For our application, these are not known a priori and thus need to be chosen adaptively.

Concerning the choice of TT ranks, we follow the adaptive strategy proposed in~\cite{steinlechner2016riemannian}.
We start by solving \eqref{completionproblem} for the smallest sensible choice of TT ranks, $\mathbf{r}=(1,\dots,1)$. Most likely, this choice will not suffice to obtain satisfactory accuracy and the  error on the test set will be relatively large.
To decrease it, the obtained solution is used as starting value for Riemannian CG applied again to \eqref{completionproblem}, but this time with the increased TT ranks $\mathbf{r}=(1,2,1,\dots,1)$ as discussed in~\cite{steinlechner2016riemannian}. See also~\cite{uschmajew2015greedy} for a greedy rank update procedure in the context of matrix completion. The described procedure is repeated by increasing cyclically every TT rank $r_\mu$. The overall algorithm stops as soon as increasing any of the TT ranks does not improve the test set error anymore or the maximal possible rank $r_{\max}$ is reached; see Algorithm~\ref{rank_adaptivity}.

\begin{algorithm}[ht]
\caption{Adaptive rank strategy}
\begin{algorithmic}[1]\label{rank_adaptivity}
\REQUIRE Data on sampling/test sets $\Omega, \Omega_C$, max rank $r_{\max}$, acceptance parameter $\rho \geq 0$
\ENSURE Completed tensor $\Xcal$ with adaptively chosen TT ranks $\mathbf{r}$, $r_\mu \leq r_{\max}$.
\STATE  $\mathcal{X}$ random tensor having TT ranks $\mathbf{r}=(1,\dots,1)$
\STATE $\Xcal \leftarrow$ result of Riemannian CG (see Section~\ref{sec-completion}) using starting guess $\Xcal$
\STATE  $\mathsf{locked}=0$
\STATE $\mu=1$
\WHILE{$\mathsf{locked}<d-1$ $\&$  $\max_\nu r_\nu < r_{\max}$}
\STATE $\Xcal_{\mathrm{new}} \leftarrow$ increase $\mu$th rank of $\Xcal$ to $(r_0,\dots,r_{\mu-1},r_{\mu}+1,r_{\mu+1},\ldots,r_{d})$
\STATE $\Xcal_{\mathrm{new}} \leftarrow$ result of Riemannian CG using starting guess $\Xcal_{\mathrm{new}}$
\IF {$(\epsilon_{{\Omega_C}}(\Xcal_{\mathrm{new}}) - \epsilon_{{\Omega_C}}(\Xcal))> -\rho$}
\STATE $\mathsf{locked}\gets \mathsf{locked} + 1$ \quad \% revert step
\ELSE 
\STATE $\mathsf{locked}\gets 0$, $\Xcal \leftarrow \Xcal_{\mathrm{new}}$ \quad \% accept step
\ENDIF
\STATE $\mu \leftarrow 1 + ( \mu \mod d-1 )$
\ENDWHILE
\end{algorithmic}
\end{algorithm}

For the adaptive choice of the sampling set $\Omega$, which has not been addressed in \cite{steinlechner2016riemannian}, we present two different strategies. The core idea is to gradually increase the size of $\Omega$ in order to improve the approximation of the tensor. Both strategies are also combined with Algorithm~\ref{rank_adaptivity} and they differ only in the measurement of the error.

The steps of the first adaptive sampling strategy are as follows.
\begin{enumerate}
\item Start with a sample set $\Omega$ of small size and a test set ${\Omega_C}$ of a certain prescribed size $|\Omega_C|$. Run Algorithm \ref{rank_adaptivity}.
\item Measure the relative error on the test set ${\Omega_C}$ and stop if the stopping criterion is satisfied. If not satisfied, add the test set ${\Omega_C}$ to the sample set $\Omega$ and create a new test set of size $|\Omega_C|$. In our applications, this corresponds to computing new option prices on the Chebyshev grid using the reference method.
\item Run again Algorithm \ref{rank_adaptivity} from line 2 to the end, by using a rank $\mathbf{r}=(1,\dots,1)$ approximation of the result from the previous step as initial guess for the CG algorithm.
\item Repeat 1-3 until a maximal sampling percentage is reached or an a priori chosen stopping criterion is satisfied.
\end{enumerate}
\begin{algorithm}[h]
\caption{Adaptive sampling strategy 1}
\begin{algorithmic}[1]\label{adaptive_sampling}
\REQUIRE Initial sampled data $P_{\Omega}  \mathcal{A}$, maximal rank $r_{\max}$ for rank adaptivity, maximal allowed size percentage $p$ (of $\Omega$)
\ENSURE Completed tensor $\Xcal$ of TT ranks $\mathbf{r}$, $r_\mu \leq r_{\max}$
\STATE Create test set ${\Omega_C}^{\mathrm{new}}$ such that $\Omega \cap {\Omega_C}^{\mathrm{new}}=\emptyset$
\STATE  Run Algorithm \ref{rank_adaptivity} with $\Omega, {\Omega_C}^{\mathrm{new}}$ and get completed tensor $\Xcal_c$.
\STATE $\mathrm{err}_{\mathrm{new}} \leftarrow \epsilon_{{\Omega_C}^{\mathrm{new}}}(\Xcal_c)$
\STATE 
\WHILE{$|\Omega|/\size(\Acal)<p$}
\STATE $\mathrm{err}_{\mathrm{old}} \leftarrow \mathrm{err}_{\mathrm{new}}$
\STATE  $\tilde{\Xcal}\leftarrow$ rank $(1,\dots,1)$ approximation of $\Xcal_c$
\STATE ${\Omega_C}^{old} \leftarrow {\Omega_C}^{\mathrm{new}}$
\STATE Create new test set ${\Omega_C}^{\mathrm{new}}$ such that ${\Omega_C}^{\mathrm{new}}\cap {\Omega_C}^{old}=\emptyset$
\STATE $\Omega \leftarrow \Omega \cup {\Omega_C}^{old}$
\STATE Run Algorithm \ref{rank_adaptivity} (from line 2 to end) with $\Omega, {\Omega_C}^{\mathrm{new}}$ and $\tilde{\Xcal}$ as starting guess. \\ Get completed tensor $\Xcal_c$ out of it.
\STATE $\mathrm{err}_{\mathrm{new}} \leftarrow \epsilon_{{\Omega_C}^{\mathrm{new}}}(\Xcal_c)$ 
\IF {stopping criterion satisfied}
\STATE Break
\ENDIF
\ENDWHILE
\STATE
\STATE $\Xcal \leftarrow \Xcal_c$
\end{algorithmic}
\end{algorithm}
The pseudo-code in Algorithm \ref{adaptive_sampling} summarizes this first strategy.

The second adaptive sampling strategy that we propose is designed in a similar way. The only difference is that the error is measured on an a priori defined fixed set $\Gamma$ and not on $\Omega_C$, which changes at each step. Therefore, this strategy follows the same steps as the first one, with the only difference that in Step 2 we measure the error on the set $\Gamma$, which has been previously defined.
The algorithm summarizing this second strategy can be obtained by replacing line 3 and line 12 in Algorithm~\ref{adaptive_sampling} with
\begin{equation*}
\mathrm{err}_{\mathrm{new}} \leftarrow \epsilon_{{\Gamma}}(\Xcal_c).
\end{equation*}

The stopping criterion of line 13 can be also defined in different ways. We choose to stop the algorithm if one of the following criteria is satisfied:

\begin{enumerate}
\item if $\mathrm{err}_{\mathrm{new}}<\mathrm{tol}$, where $\mathrm{tol}$ is a prescribed tolerance;
\item if $|\mathrm{err}_{\mathrm{new}} - \mathrm{err}_{\mathrm{old}}| <  \mathrm{tol}' $ where $\mathrm{tol}'$ is a prescribed tolerance;
\item if $\exists \mu$ such that $r_{\mu}(\Xcal_c)==r_{\max}$.
\end{enumerate}
The first criterion allows us to stop as soon as the error goes below a certain level, the second stops the algorithm whenever the error stagnates and the last one when the TT rank has reached the maximal allowed rank at least in one mode $\mu$.

We test the new adaptive sampling strategies on a problem with known solution in the next section.
\subsubsection{Numerical test for adaptive sampling strategies}
\begin{figure}[ht]
\centering
\includegraphics[width=0.81\textwidth]{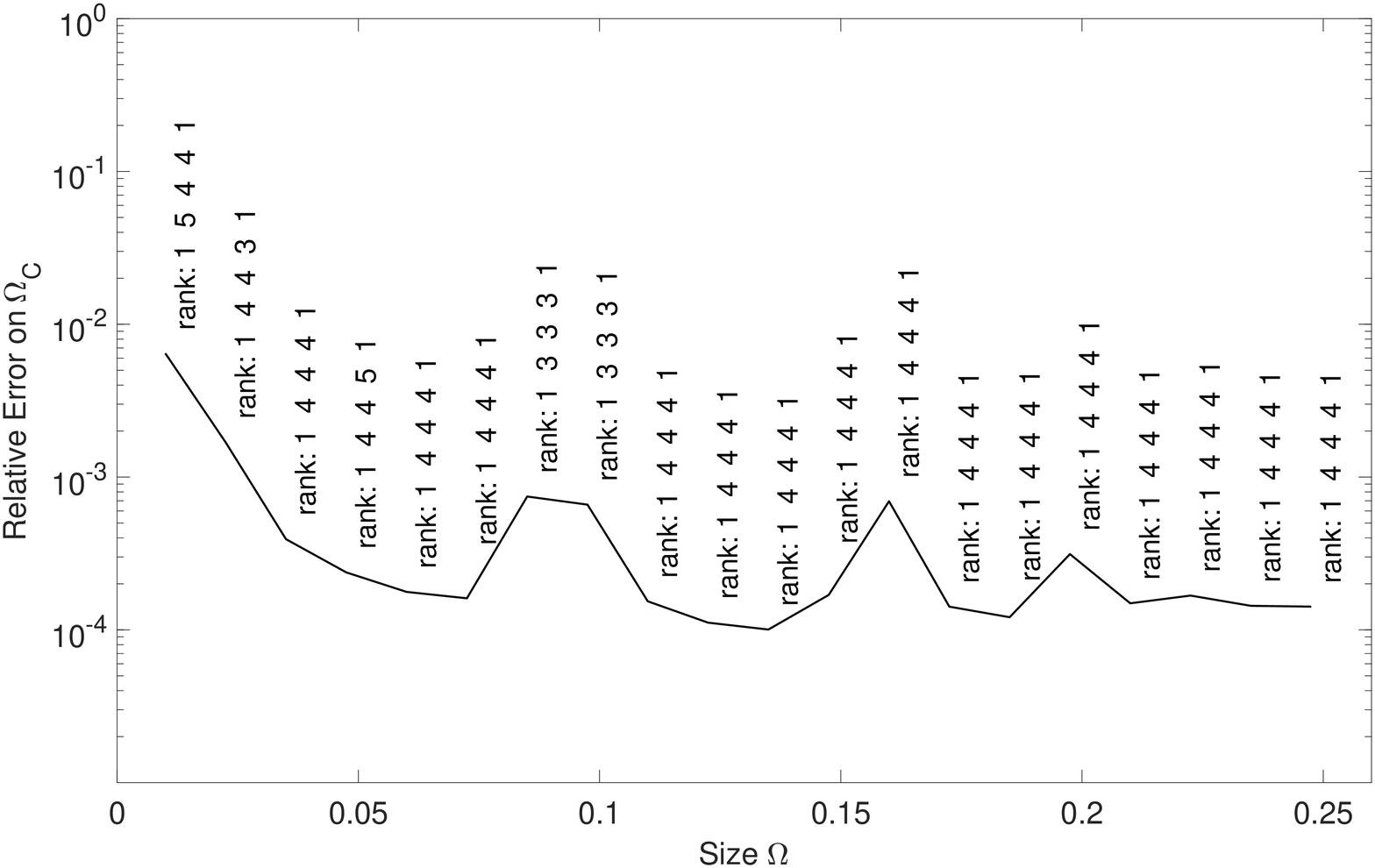}
\caption{Relative error on varying test sets $\Omega_C$ for different sampling set sizes in adaptive sampling strategy 1. \label{fig_adaptivesampling1}}
\includegraphics[width=0.81\textwidth]{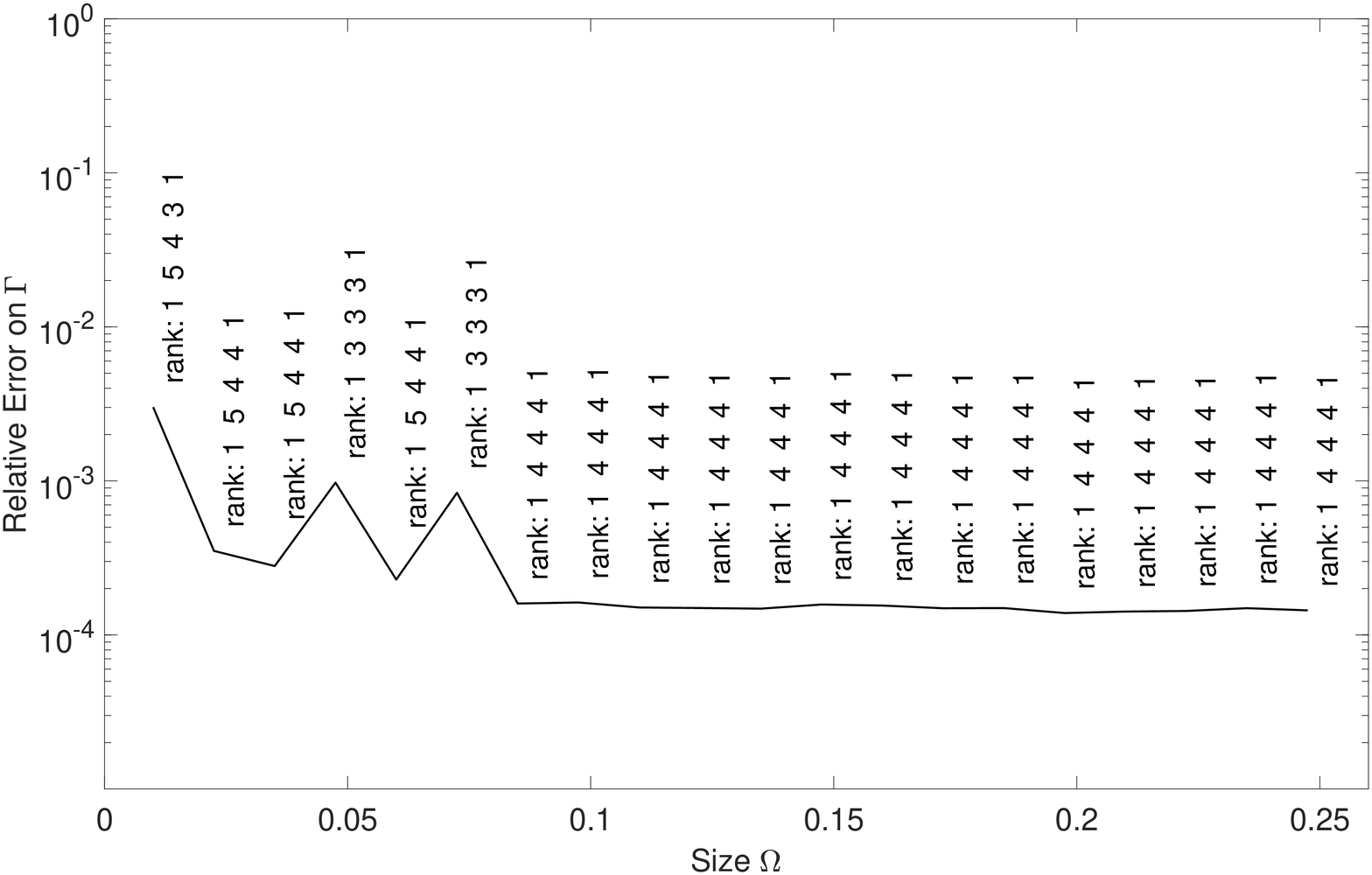}
\caption{Relative error on set $\Gamma$ for different sampling set sizes in adaptive sampling strategy 2. \label{fig_adaptivesampling2}}
\end{figure}

We consider the problem of Chapter 5.4.2 in \cite{steinlechner2016riemannian} and we apply our adaptive sampling strategies to it in order to compare them and to investigate their advantages and disadvantages. We expect a similar performance of both strategies in terms of accuracy and compression. In this numerical example, as well as in the rest of the paper, we choose $\| \cdot \|$ to be the 2-norm and $\delta=10^{-4}$ in \eqref{stoppingCG}. The problem consists of discretizing the function 
\begin{equation*}
f:[0,1]^4 \to \mathbb{R}, \quad f(\mathbf{x})=\exp(-\| \mathbf{x} \|)
\end{equation*}
using $n=20$ equally spaced discretization points on $[0,1]$ in each mode. We aim at reconstructing the tensor containing the function values in the grid. In Algorithm~\ref{adaptive_sampling} we set the maximum rank to $\mathbf{r}_{\max}=(1,7,7,7,1)$ and we start with an initial sampling set $\Omega$ satisfying $|\Omega|/n^4=0.01$. Moreover, we set the acceptance parameter $\rho$ of Algorithm \ref{rank_adaptivity} to $\rho = 10^{-4}$. In order to analyze the behavior of the error, we do not impose any stopping criterion, but we let our adaptive sampling strategies run until $|\Omega|/\size(\Acal)>0.25$. The size of each $\Omega_C$ is set to $2000$ and $|\Gamma|=3000$ for the second strategy. Figures \ref{fig_adaptivesampling1} and \ref{fig_adaptivesampling2} show the results for the two different strategies. 

First, we observe that both strategies eventually reach the same accuracy and the same final TT ranks, which makes both of them valid. We observe an oscillatory behavior in Figure~\ref{fig_adaptivesampling1}. This non-smooth decay can be expected since in each step the error is measured on a different test set $\Omega_C$. We observe that the amplitude of the oscillations becomes smaller as $|\Omega|$ increases. This indicates an error stagnation over the whole tensor which cannot be improved by enlarging $\Omega_C$ further. On the other hand, the error in the second strategy behaves almost monotonically and stagnates much earlier than in the previous case. This is due to the fact that we measure it on the fixed set $\Gamma$. In practice, the earlier error stagnation of the second strategy is preferable as it triggers the stopping criterion 2. However, the second strategy has the disadvantage of the initial additional cost of evaluating the tensor in the set $\Gamma$. In our numerical experiments in Section~\ref{section3} we choose the first strategy, which turned out to be more favorable since the stopping criterion 1 was triggered.

\subsection{Combined methodology}\label{sec-combined methodology}

We are now in the position to combine the concepts and the algorithms in order to develop an efficient procedure for high-dimensional tensorized Chebyshev interpolation.

We would like to price options that depend on a vector $\p=(p_1,\cdots, p_d)$ of $d$ varying parameters. It is reasonable to assume that every combination of parameters $\p$ belongs to a compact hyper-rectangular $[\underline{p}_1,\overline{p}_1] \times [\underline{p}_2,\overline{p}_2] \times \cdots \times [\underline{p}_d,\overline{p}_d]$. For example, if time-to-maturity $T$ belongs to the set of varying parameters, we can assume that $T \in [0.05, 2]$; similarly for the other payoff or model parameters. The combined methodology consists of two phases: offline phase and online phase, as already introduced in~\cite{gass2016chebyshev}. 
\subsubsection{Offline phase - Computation of $\Pcal$}
The offline phase starts by performing following operations:
\begin{enumerate}
\item Fix an interpolation order $\overline{\mathbf{n}}=(n_1, \dots, n_d)$ and compute the entries of the tensor $\Pcal$ (as defined in \eqref{coeff}) from an a priori chosen subset $\Omega$ of Chebyshev nodes, using the reference pricing technique.
\item Apply tensor completion with adaptive sampling strategy (Algorithm \ref{adaptive_sampling}) in order to get a low-rank approximation of the tensor $\Pcal$ in TT format. 
\end{enumerate}

For simplicity, we denote the obtained low-rank approximation of $\Pcal$ again by $\Pcal$. In the last step of the offline phase we construct the interpolation coefficients, defined in \eqref{coeff}. We denote the tensor of coefficients by $\mathcal{C} \in \mathbb{R}^{(n_1+1) \times (n_2+1) \times \cdots \times (n_d+1)}$. Its entries are therefore given by (adjusting the ordering according the Sections \ref{chebyshevPOP} and \ref{sec-ttformat}) 
\begin{equation}
\mathcal{C}(i_1, i_2, \cdots, i_d) = c_{i_1-1, i_2-1, \cdots, i_d-1}, \label{Ctensor} 
\end{equation}
for $i_j=1,\cdots, n_j+1$ and $j=1, \cdots, d$. The tensor $\Ccal$ can be efficiently computed in TT format, as explained in the following subsection.

\subsubsection{Offline phase - Efficient computation of $\mathcal{C}$}\label{computationC}
In order to explain the algorithm we first consider the simple case $d=1$. In this case $\Pcal$ and $\Ccal$ are in $ \mathbb{R}^{(n_1+1) \times 1}$, where $n_1$ is the chosen interpolation order. The entries of $\Ccal$ are given by
\begin{equation*}
\Ccal(j+1)=\frac{2 ^{\mathbbm{1}_{n_1>j >0}}}{n_1}  \sideset{}{''}\sum_{k=0}^{n_1} \Pcal (k+1)\cos \big(j \pi \frac{k}{n_1}   \big), \quad j=0, \cdots, n_1,
\end{equation*}
so that the whole vector $\Ccal$ can be computed via the matrix-vector multiplication
\begin{equation}\label{fft}
\Ccal = 
\frac{2}{n_1}
\begin{pmatrix}
\frac{1}{4} & \frac{1}{2} & \hdots& \frac{1}{2}& \frac{1}{4}     \\  
\frac{1}{2} & \cos( \frac{\pi}{n_1})& \hdots & \cos(\frac{\pi (n_1-1)}{n_1})& \frac{1}{2} \cos(\pi) \\
\vdots &\vdots& \ddots& \vdots& \vdots\\
\frac{1}{2} & \cos( \frac{\pi (n_1-1)}{n_1})& \hdots & \cos(\frac{\pi (n_1-1)^2}{n_1})& \frac{1}{2} \cos(\pi (n_1-1)) \\
\frac{1}{4} & \frac{1}{2} \cos(\pi) & \hdots& \frac{1}{2} \cos(\pi (n_1-1))& \frac{1}{4}   \cos(\pi n_1)   \\  
\end{pmatrix}
\Pcal,
\end{equation}
and we denote by $F_{n_1} \in \mathbb{R}^{(n_1+1) \times (n_1+1)}$ the matrix multiplying $\Pcal$ in \eqref{fft}.

For a general dimension $d >1$, the same reasoning can be applied and the tensor $\Ccal$ of interpolation coefficients can be computed by sub-sequentially multiplying $\Pcal$ with $F_{n_i}$ ($i=1, \cdots, d$) via the mode-$\mu$ multiplication, defined in Section \ref{sec-ttformat}. The final procedure for an efficient computation of $\Ccal$ is given in Algorithm \ref{computation_Ccal}.
\begin{algorithm}[ht]
\caption{Efficient computation of $\Ccal$}
\begin{algorithmic}[1]\label{computation_Ccal}
\REQUIRE Tensor $\Pcal$ in TT format containing option prices in the Chebyshev grid
\ENSURE Tensor $\Ccal$ as defined in \eqref{Ctensor}, in TT format
\STATE Compute $F_{n_1}$ as in \eqref{fft}.
\STATE $\Ccal \leftarrow \Pcal \times_1 F_{n_1}$
\FOR {$m=2,\dots, d$}
\STATE Compute $F_{n_m}$
\STATE $\Ccal \leftarrow \Ccal \times_m F_{n_m}$.
\ENDFOR
\end{algorithmic}
\end{algorithm}

Note that if $n_1= \cdots =n_d \bydef n$ (as for example in our numerical experiments in Section \ref{section3}), Algorithm \ref{computation_Ccal} can be further simplified by computing the matrix $F_n$ only once. The particular structure of the matrices $F_{n_i}$ allows us to apply a Fast-Fourier-Transform based algorithm which computes each mode multiplication in $O(r^2 n \log(n))$ (instead of $O(r^2 n^2$) as mentioned in Section \ref{section3}). Therefore, the total complexity for computing $\Ccal$ is $O(dnr^2\log(n))$.

The offline phase can be finally completed by performing the step 
\begin{enumerate}
\setcounter{enumi}{2}
\item Construct the tensor $\mathcal{C}$ as explained in Algorithm \ref{computation_Ccal}. \label{constructionC}
\end{enumerate}
\subsubsection{Online phase}
\begin{figure}[ht]
\centering
\includegraphics[width=0.8\textwidth]{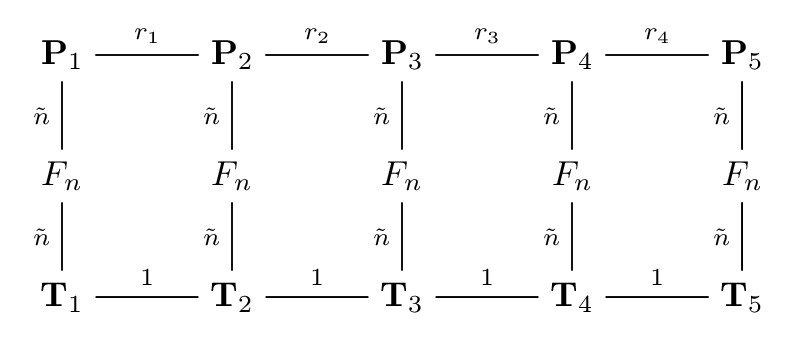}
\caption{Tensor network diagram representing the whole interpolation procedure as in \eqref{intprice} and \eqref{coeff}, for $d=5$ in TT format. Note that $\tilde{n}:= n+1$. \label{intertt}}
\end{figure}

Once we have stored $\Ccal$ in TT format, we can use it to compute every option price via interpolation during the online phase. For any particular choice of parameters $\p$, we first perform the step
\begin{enumerate}
\setcounter{enumi}{3}
\item Evaluate the Chebyshev tensor basis \eqref{cheb-basis} in $\p$.
\end{enumerate}
This step returns a tensor $\mathcal{T}_\p \in \mathbb{R}^{(n_1+1) \times (n_2+1) \times \cdots \times (n_d+1)}$ of TT rank $(1,\cdots, 1)$, that we store in TT format. 
The interpolated price, defined in \eqref{intprice}, can now be rewritten as the inner product
\begin{equation}
I_{\overline{\mathbf{n}}}(\mathsf{Price}^{(\cdot)})(\p) = \langle \mathcal{C}, \mathcal{T}_\p \rangle. \label{Pricetensor}
\end{equation}
The final step of our combined methodology is then defined as
\begin{enumerate}
\setcounter{enumi}{4}
\item Compute the interpolated price \eqref{Pricetensor} in TT format as in \eqref{innerproduct}.\label{computeprice}
\end{enumerate}
If we consider a fixed interpolation order $n$ in each dimension and if the TT ranks of $\Pcal$ and $\Ccal$ are approximately $r$, then the total cost for performing both Step~\ref{constructionC} and Step~\ref{computeprice} is given by $O(dnr^2+dnr^2\log(n))$. These two steps are represented via a tensor network diagram in Figure \ref{intertt} (for $d=5$), where we denoted by $\mathbf{P}_i$ the core tensors of $\Pcal$ and by $\mathbf{T}_i$ the ones of $\mathcal{T}_\p$.

Finally, we summarize our complete methodology in Algorithm \ref{methodology}.
\begin{algorithm}[ht]
\caption{Combined methodology for Chebyshev interpolation in parametric option pricing}
\begin{algorithmic}[1]\label{methodology}
\REQUIRE Interpolation order $\overline{\mathbf{n}}$, subset $\Omega$ of total Chebyshev points, set $\Pi$ of parameters $\p$ for which we want to compute option prices
\ENSURE Interpolated option prices for parameters $\p\in \Pi$
\STATE  \% \sl Offline phase \rm
\STATE  Compute option prices using reference method in the subset $\Omega$ of Chebyshev points
\STATE  Construct $\Pcal$ using tensor completion in TT format (Algorithm \ref{adaptive_sampling})
\STATE  Construct tensor $\mathcal{C}$ as in Section \ref{computationC}
\STATE
\STATE  \% \sl Computation of option prices - Online phase \rm
\FOR {$\p \in \Pi$}
\STATE  Evaluate the Chebyshev tensor basis $\mathcal{T}_\p$
\STATE  Compute interpolated price \eqref{Pricetensor}
\ENDFOR  
\end{algorithmic}
\end{algorithm}

In the next section we see how this combined methodology performs on concrete examples.

\section{Financial applications and numerical experiments}\label{section3}
Putting the new approach to test,
we implement the method described in Section~\ref{section2} for two different types of applications. In the first one, we tackle computational intense option pricing methods in a parametric model.  We treat option prices as functions in the parameter space which consists of model and option parameters. We then approximate the price function by Chebyshev interpolation in the parameter space. This approach has been successfully tested in cases where the parameter space is low-dimensional.
In various applications, several varying parameters are of interest. If the interpolation is even efficient in the full parameter space, it is indeed a new pricing methodology. Here, we combine Chebyshev interpolation and low-rank approximation to cope with higher dimensionality in the parameter space. Already for pricing single asset options, it is promising to tackle medium and high-dimensional parameters spaces in this approach.
As a generic example, we choose to approximate American put option prices in the Heston model with the varying parameters $K, \rho,\sigma,\kappa$ and $\theta$. It turns out that the computational complexity reduces significantly in this case.
 
As second type of application we examine the interpolation of basket option prices in the $d$-variate Black-Scholes model as function of the initial stock prices. This is a prototypical example for the computation of generalized conditional moments of high-dimensional Markov processes.

All algorithms have been implemented in {\sc Matlab} and run on a standard laptop (Intel Core i7, 2 cores, 256kB/4MB L2/L3 cache). In order to deal with tensors, we used the toolboxes \cite{oseledets2011} by Oseledets and \cite{brett2006algorithm,TTB_Software}, while for the completion algorithm we used the TT completion toolbox described in \cite{kressner2014low, kressner2016preconditioned, steinlechner2016riemannian, steinlechner2016thesis}. Note that in this toolbox the most expensive steps have been implemented in C using the {\tt Mex}-function capabilities of {\sc Matlab}.

\subsection{Pricing American options in Heston's model}\label{sectionHeston}
We consider pricing single asset American put options in the Heston model. As introduced by Heston in~\cite{heston1993closed}, the price dynamics of the financial asset under the risk neutral measure are given by 
\begin{equation*}
dS_t = r S_tdt+\sqrt{v_t}S_t dW_t^1,
\end{equation*}
where  the square of the volatility $v_t$ is modeled by the square root process
\begin{equation*}
dv_t = \kappa (\theta-v_t)dt+\sigma \sqrt{v_t} dW_t^2.
\end{equation*}
Here, the two Brownian motions $W^1$ and $W^2$ are correlated with correlation parameter $\rho$, mean-reversion rate $\kappa >0$, long-term mean $\theta >0$,  volatility of the variance $\sigma >0$ and, finally, fixed and deterministic continuously compounding interest rate $r$.

The price of an American option at time $t<T$, maturing at $T$, with initial underlying price $s\geq 0$ and initial volatility $v \geq 0$ is given by 
\begin{equation}\label{american_price}
\mathsf{Price} =  \sup_{t<\tau<T} \mathbb{E}[e^{-r\tau} f(S_\tau)|S_t=s, v_t=v],
\end{equation}
where the $\sup$ is taken over all stopping times $\tau$ in $[t, T]$. Here, $f$ denotes the payoff function of the European put option, i.e. 
\begin{equation*}
f(x)=(K-x)^+,
\end{equation*}
where $K$ denotes the strike price.

It is well-known (see e.g.~\cite{duffy2006finite}) that the price \eqref{american_price} of the American option satisfies the following partial differential complementarity problem (PDCP):
\begin{align}
\begin{split}\label{PDCP}
\begin{cases}
\partial_t \mathsf{Price} \geq \mathcal{G} \mathsf{Price}\\
\mathsf{Price} \geq f\\
(\mathsf{Price} - f)(\partial_t \mathsf{Price}-\mathcal{G} \mathsf{Price})=0,
\end{cases}
\end{split}
\end{align} 
where $\mathcal{G}$ is the infinitesimal generator of $(s,v)$ in the Heston model, defined as
\begin{equation*}
\mathcal{G}g(s,v) = \frac{1}{2} s^2v \partial^2_{ss} g +\rho \sigma s v \partial^2_{sv} g +\frac{1}{2} \sigma^2 v \partial^2_{vv} g+rs
\partial_s g+\kappa(\theta-v) \partial_v g -r g.
\end{equation*}
The problem \eqref{PDCP} has been well studied in the literature and different pricing algorithms have been developed so far. In our example we consider, as reference method for our combined methodology, the pricing algorithm explained in~\cite{haentjens2015adi}. More precisely, the authors propose different schemes for the time discretization and we consider the Hundsdorfer Verwer - Ikonen Toivanen (HV-IT) scheme, explained at page 219 of~\cite{haentjens2015adi}.

Solving the discretized PDCP yields an approximate price for all values of $S_0, v_0$ and $T$ in each grid point of the pre-specified domain. For many applications we would like to have the solution at hand for other parameters (as well). In calibration, for instance, we observe $S_0$ and $r$, and one could estimate $v_0$ from historical stock price data. Then the calibration problem reduces to fitting the parameters $(K, \rho,\sigma,\kappa,\theta)$ to the observed option price data. To do so one needs to solve an optimization problem where prices need to be computed for large sets of parameters $(K, \rho,\sigma,\kappa,\theta,T)$. Since the price for different maturities can be obtained by rescaling $\kappa$ and $\sigma$, effectively we need the prices for combinations of the parameters $K, \rho,\sigma,\kappa$ and $\theta$. This motivates the following set up, where we fix the model and payoff parameters 
\begin{equation*}
S_0=2, \quad v_0=0.0175, \quad r=0.1, \quad T=0.25,
\end{equation*} 
and we let vary the five parameters
\begin{equation*}
(K, \rho,\sigma,\kappa,\theta) \in [2;4] \times [-1;1] \times [0.2;0.5] \times [1;2] \times [0.05;0.2]
\end{equation*}
in their corresponding domain.

In order to compute the reference prices we consider $50$ equidistant spatial grid points in both directions $s$ and $v$ with $s_{\min}=0,s_{\max}=5,v_{\min}=0,v_{\max}=1$, $40$ time steps and the Crank-Nicholson time stepping scheme.

We start by performing the offline phase of Algorithm \ref{methodology}. We consider an interpolation order $n_1= \cdots=n_5\bydef n=10$ in each direction and we construct the tensor $\Pcal$ by tensor completion as explained in Section \ref{sec-completion}. We apply the first adaptive sampling strategy as in Algorithm \ref{adaptive_sampling}. We choose the completion parameters as
\begin{equation*}
\rho = 0, \quad tol=10^{-3}, \quad tol'=10^{-8}, \quad r_{\max}=10, \quad |\Omega|=805, \quad |\Omega_C|=805, \quad p=0.2.
\end{equation*}
For this particular example, we were also able to explicitly construct the full tensor (in more than 1 hour and 40 minutes!). In Table \ref{completionAmerican} we show the size of the final set $\Omega$ (first column), the relative error of the completed tensor on the last $\Omega_c^{new}$ (second column), the relative error between the obtained completed tensor and the full one (third column), the runtime of the completion, Algorithm~\ref{adaptive_sampling}, in seconds (fourth column), the TT-rank of $\Pcal$ (fifth column), the storage needed to save $\Pcal$ in TT format, denoted by store(TT) and measured in bytes (sixth column) and finally, the storage needed to save the full tensor, denoted by store(full) and again measured in bytes. {\sc Matlab} requires $8$ bytes to store a floating-point number of type double, which gives us the formula store(full)$=8 \cdot(n+1)^d$ for the storage of the full tensor and store(TT)$=8 \cdot(n+1)(r_1r_2+\cdots+r_{d-2}r_{d-1})+8 \cdot(n+1)(r_1+r_{d-1})$ for the storage of the tensor in TT format, see~\cite{oseledets2011}.

Table \ref{completionAmerican} shows that a sample set of $5\%$ is sufficient for the algorithm to reach the prescribed accuracy. Furthermore, the relative error of the completed tensor $\Pcal$ in the 2-norm over the last test sample parameter space $\Omega^{new}_c$ and the relative error over the full tensor, i.e. over all Chebyshev nodes, is only in the $6$th digit. This is one order of magnitude smaller than the relative error on the full $\Pcal$. This is a good indication that the approach can be extended to more complex cases, where the computation of the full tensor $\mathcal{P}$ is not feasible any more (see Section \ref{numerical exp BS}). The completion time was about $6$ minutes. Finally, the rank properties together with its storage reduction of a factor of $115$ confirm the low-rank structure of the problem.

\begin{table}[ht]
\centering
    \begin{tabular}{cccc}
    final $|\Omega|$ & rel err on last $\Omega^{new}_c$ & rel err on full $\Pcal$& completion time (s) \\  \hline \rule{0pt}{1.0\normalbaselineskip}
     8050 (5 \%)& $2.56 \cdot 10^{-5}$& $2.75 \cdot 10^{-5}$ &366.12  \vspace{0.3cm}
\end{tabular}
    \begin{tabular}{ccc}
    $\ranktt(\Pcal)$ & store(TT) (bytes) & store(full) (bytes)\\ \hline \rule{0pt}{1.0\normalbaselineskip}
     $(1,5,8,6,5,1)$ &11264 & 1288408
\end{tabular}
\caption{Completion results on $\Pcal$ for the parametric American put option pricing problem in the Heston model. \label{completionAmerican}}
\end{table}

For constructing the tensor $\mathcal{C}$ (last step of the offline phase) we applied Algorithm \ref{computation_Ccal} and the computation time was $0.0037$ seconds, which is negligible compared to the completion time. Hence, almost all the computation time in the offline phase is spent in the construction of the tensor $\Pcal$.

Next, we compute American put option prices for the online phase in both ways using our methodology and the reference algorithm. We compute $243$ prices with random model parameters uniformly drawn from the reference set $[2;4] \times [-1;1] \times [0.2;0.5] \times [1;2] \times [0.05;0.2]$. We measure the maximal absolute error over the computed options prices, i.e. we report the quantity 
\begin{equation*}
\max(|P_{\text{Int}}-P_{\text{Ref}}|),
\end{equation*}
where $P_{\text{Int}}$ is a vector containing all interpolated prices for the different choices of model parameters; analogously is $P_{\text{Ref}}$ for the reference method.
In Table \ref{pricesAmerican} we also report the computation time for computing one single option price for both methods. One can notice that the online phase of the interpolation compared to the reference method accelerates the procedure by a factor of $75$. The accuracy of the reference method is reported in part C of Figure 1 in \cite{haentjens2015adi} for one specific parameter set to be of the order $10^{-3}$ in the maximum norm. The interpolation error is one order smaller, making the new procedure at least as accurate as the reference method. Therefore, we can conclude that the methodology strongly outperforms the reference method in the online phase while keeping the same accuracy. 

We would like to emphasize that this approach can be further extended to an interpolation in the full set of parameters $(S_0, v_0, r, T, K, \rho,\sigma,\kappa,\theta)$. Since then the offline phase needs to be performed only once, this would result in a new pricing method. Here, in the offline phase one could explore the fact that the PDCP solver returns the price for all $(S_0,v_0,T)$ in the grid to make the sampling steps more efficient. This opens up an interesting topic for future research.

\begin{table}[ht]
\centering
    \begin{tabular}{ccc} 
     time reference method (s) & time interpolation (s)& max abs error\\  \hline \rule{0pt}{1.0\normalbaselineskip}
  $3.65 \cdot 10^{-2}$  & $4.89 \cdot 10^{-4}$& $1.95 \cdot 10^{-4}$\\ 
    \end{tabular}
\caption{Results on American put option pricing via combined methodology and reference method.
\label{pricesAmerican}}
\end{table}
\subsection{Basket options in multivariate Black-Scholes model}\label{numerical exp BS}
In the $d$-variate Black-Scholes model with $d$ assets $S^1, \cdots, S^d$, the risk neutral dynamics are given by
\begin{equation}\label{SDE_BS}
dS^i_t = r S_t^i dt+\sigma_i dW^i_t,
\end{equation} 
where $r$ is a fixed deterministic interest rate, $(\sigma_1, \cdots, \sigma_d)$ is the vector of volatilities and $(W^1, \cdots, W^d)$ is a vector of correlated Brownian motions with correlation matrix $\Sigma$. The solution to \eqref{SDE_BS} is given by
\begin{equation*}
S_t^i = S_0^i \exp\big( (r-\frac{\sigma_i^2}{2})t + \sigma_i W_t^i \big).
\end{equation*}
In this section we apply the new methodology in order to price basket options with payoff function $f :\mathbb{R}^d \to \mathbb{R}$ defined as 
\begin{equation*}
f(\mathbf{x}) \defby \Big (\sum_{n=1}^d w_n x_n - K\Big )^+,
\end{equation*}
where $K$ is the strike and $(w_1, \cdots, w_d)$ is a vector of weights satisfying $\sum_{n=1}^d w_n=1$. The risk neutral price at time $t=0$ of the basket option with maturity $T$ is, as usual, given by 
\begin{equation}\label{basketprice}
\mathsf{Price} = e^{-rT} \mathbb{E}[f(\mathbf{S_T})].
\end{equation}
From now on, we consider the parameters $r, \sigma_i$ $(i=1, \cdots, d)$ and the correlation matrix $\Sigma$ to be fixed, and we let the vector $\mathbf{S_0} \in \mathbb{R}^d$ of initial asset prices be the varying parameter. The reference pricing algorithm will be of Monte Carlo (MC) type combined with a variance reduction technique. In particular, we use the control variates method presented in \cite{glasserman2004monte}, where the control variate is given by 
\begin{equation*}
Y \bydef \Big(\exp \big (\sum_{i=1}^d \omega_i \log(S_T^i) \big)-K\Big)^+.
\end{equation*}
Since the only varying parameter is the vector of initial asset prices, it is very convenient to split the Monte Carlo simulation in two parts in order to make the completion more efficient. More precisely, in a pre-computation phase (Algorithm \ref{SimCorrGBM}) we simulate a certain number of realizations (e.g. $10^4$) of 
\begin{equation*}
\exp\big( (r-\frac{\sigma_i^2}{2})T + \sigma_i W_T^i \big), \quad \text{for } i=1, \cdots, d,
\end{equation*}
and in a second moment we multiply the vector $\mathbf{S_0}$ (for all required parameter combinations) with all the realizations and we compute the Monte Carlo price by applying the chosen variance reduction technique (Algorithm \ref{BasketNdSim}). In order to generate the correlated random variables $W_T^i$, we use the Cholesky factorization of the correlation matrix, which is then multiplied by a vector of independently generated standard normal distributed random variates.
Note that $\circ$ in Algorithm \ref{BasketNdSim} represents the Hadamard (component-wise) product between vectors.
\begin{algorithm}[ht]
\caption{Simulation of correlated geometric Brownian motions}
\begin{algorithmic}[1]\label{SimCorrGBM}
\REQUIRE Model and payoff parameters $\mathbf{\sigma}, \Sigma, T, r$; number of simulations $NumberSim$.
\ENSURE Matrix $M \in \mathbb{R}^{NumberSim \times d}$ containing simulated random variables.
\STATE $L \leftarrow$ Cholesky factor of $\Sigma$
\STATE $M \leftarrow$ zeros(NumberSim, d)
\FOR {iSim = 1:NumberSim}
\STATE $\epsilon \leftarrow$ Generate a vector of $d$ independent standard normal variates
\STATE $x \leftarrow L \epsilon$
\FOR {iStock = 1:d}
\STATE $M(iSim, iStock) \leftarrow \exp((r-\frac{\sigma(iStock)^2}{2})T+\sigma(iStock) x(iStock)\sqrt{T})$ 
\ENDFOR
\ENDFOR
\end{algorithmic}
\end{algorithm}
\begin{algorithm}[ht]
\caption{Computation of basket options using MC with control variate technique}
\begin{algorithmic}[1]\label{BasketNdSim}
\REQUIRE Matrix M from Algorithm \ref{SimCorrGBM}, $\mathbf{S}_0$, strike $K$, vector of weights $\omega$, $r$, $T$
\ENSURE Basket option price \eqref{basketprice}
\STATE $payoff \leftarrow$ zeros(NumberSim,1)
\STATE $control \leftarrow$ zeros(NumberSim,1)
\FOR {iSim = 1: NumberSim}
\STATE $R \leftarrow iSim$-th row of $M$
\STATE $S \leftarrow S_0 \circ R^T $
\STATE $payoff(iSim) \leftarrow (\sum_{i=1}^d \omega_i S_i - K)^+$
\STATE $control(iSim) \leftarrow (\exp(\sum_{i=1}^d \omega_i \log(S_i))-K)^+$
\ENDFOR
\STATE Compute mean $\mu_Y$ of $Y$ as explained in \cite{glasserman2004monte}
\STATE $sum \leftarrow payoff -(control -\mu_Y)$
\STATE Compute mean $\mu$ of $sum$
\STATE $Price \leftarrow \exp(-rT)\mu$
\end{algorithmic}
\end{algorithm}

Algorithm~\ref{SimCorrGBM} is executed at the beginning of the whole procedure and Algorithm~\ref{BasketNdSim} whenever needed in later stages. The advantage of splitting the MC algorithm is twofold. Firstly, it supports a considerable gain in efficiency in the performance of the completion algorithm: When we adaptively increment the sampling set $\Omega$ (which consists of sampling Chebyshev nodes in $\mathbf{S}_0$) in Algorithm~\ref{adaptive_sampling}, we need to compute new prices in the Chebyshev grid, which can be done by using Algorithm~\ref{BasketNdSim} only. The second advantage regards the analysis of the methodology and the completion accuracy: Since we use the same set of simulations for every Chebyshev price, the MC simulation does not introduce any further error to the completion. Moreover, we will see in Section~\ref{analysisP} that this splitting procedure allows for a qualitative analysis of the rank structure of $\Pcal$.

Next, we perform numerical experiments for different settings of model parameters, first for uncorrelated then for correlated assets.

\subsubsection{Basket options of uncorrelated assets}
In this example we consider the special case of uncorrelated assets. We investigate the performance of the proposed method for two different interpolation orders $n_1=\cdots=n_d\bydef n=4$ and $n_1=\cdots=n_d\bydef n=6$. We apply the combined methodology (Algorithm \ref{methodology}) to portfolios consisting of $d \in \{5,10,15,20,25\}$ assets. The set of fixed parameters is given by
\begin{align*}
T=0.25, \quad K=1,\quad r=0, \quad \sigma_i = 0.2 \enskip \forall i, \quad \Sigma= I_d, \quad \omega_i = \frac{1}{d} \enskip \forall i,
\end{align*}
where $I_d$ denotes the $d \times d$ identity matrix. We let $\mathbf{S}_0$ vary in the hyper-rectangular 
\begin{equation*}
[1; 1.5]^d,
\end{equation*}
so that we consider ITM options and ATM options as well.

For each value of $d$, we start by performing Algorithm~\ref{SimCorrGBM} with $NumberSim = 10^3$ for $n=4$ and with $NumberSim = 10^4$ for $n=6$. In a second moment we construct the tensor $\Pcal$ by applying the tensor completion with the adaptive sampling strategy of Algorithm~\ref{adaptive_sampling} (first strategy). Table~\ref{compl_param} shows the completion parameters for each value of $d$ and each interpolation order.
The results of the tensor completions are displayed in Table~\ref{compl_results}. As in the previous subsection, we report the final size of the set $\Omega$, the relative error measured on the last set $\Omega_C^{new}$, the completion time and the memory needed to store both the obtained tensor in TT format and the full tensor. For the TT ranks of the completed tensor, we do not report the full tuple $(r_0, \cdots, r_d)$ (see Definition \eqref{ttrankdef}) but only the quantity $\max_{\mu \in \{0, \cdots, d\}}r_{\mu}$.

\begin{table}[ht]
\centering
    \begin{tabular}{ccccccccc}
  &  $d$ & $\rho$ & $tol$ & $tol'$ &$ r_{\max}$ & initial $|\Omega|$  & $|\Omega_C|$ & $p$\\ \hline  \rule{0pt}{1.0\normalbaselineskip}
   $n=4$ &5 &   0  &    $10^{-2}$     &       $10^{-8}$             &     $5$             &               31            &         31           &  $10^{-1}$ \\
    &10 &   0   & $10^{-2}$     &    $10^{-8}$          &          $5 $     &                      78     &                78    &  $10^{-2}$ \\
    &15 &    0  & $10^{-2}$     &    $10^{-8}$          &          $5 $      &                     214      &               214     & $10^{-5}$  \\
    &20 &    0  & $10^{-2}$      &   $10^{-8}$         &        $ 5$      &                     763   &               763    &  $10^{-8} $\\
    &25 &   0   &  $10^{-2}$      & $10^{-8}$           &       $ 5$       &                     2086      &               2086     & $10^{-11} $ \\
    \hline \rule{0pt}{1.0\normalbaselineskip}
    $n=6$ &5 &   0  &    $10^{-3}$     &       $10^{-8}$             &     $7$             &               17            &         17           &  $10^{-1}$ \\
    &10 &   0   & $10^{-3}$     &    $10^{-8}$          &          $7  $     &                      282     &                141    &  $10^{-3}$ \\
    &15 &    0  & $10^{-3}$     &    $10^{-8}$          &          $7 $      &                     475      &               475     & $10^{-6}$  \\
    &20 &    0  & $10^{-3}$      &   $10^{-8}$         &        $ 7 $      &                     798      &               798     &  $10^{-10} $\\
    &25 &   0   &  $10^{-3}$      & $10^{-8}$           &       $ 7 $       &                     1341      &               1341     & $10^{-15} $ \\
    \end{tabular}
\caption{Completion parameters for constructing $\Pcal$. Case of uncorrelated assets. \label{compl_param}}
\end{table}

\begin{table}[ht]
\centering
    \begin{tabular}{ccccc}
    &$d$ & final $|\Omega|$& rel err on last $\Omega_C^{new}$ & completion time (s) \\ \hline \rule{0pt}{1.0\normalbaselineskip}
    $n=4$&5 & 124  &$3.42 \cdot 10^{-3}$& 9.90    \\
    &10 & 546& $2.54 \cdot 10^{-6}$&  67.44  \\
    &15 & 1712 & $3.55 \cdot 10^{-8}$&  171.14    \\
    &20 & 2289 & $5.03 \cdot 10^{-8}$&   193.90     \\
    &25 &  4172 & $3.96 \cdot 10^{-9}$&  226.38 \\
    \hline \rule{0pt}{1.0\normalbaselineskip}
    $n=6$&5 & 204  &$2.40 \cdot 10^{-4}$& 52.55     \\
    &10 & 987 & $1.20 \cdot 10^{-6}$&  198.27   \\
    &15 & 1900 & $2.28 \cdot 10^{-7}$&  429.39    \\
    &20 & 3192 & $2.97 \cdot 10^{-7}$&   732.49     \\
    &25 &  4023 & $1.35 \cdot 10^{-7}$&  999.25  
    \end{tabular}
    
    \vspace{0.5 cm}
    
        \begin{tabular}{ccccc}
     &$d$ &  max $r_{\mu}$ reached & store(TT) (bytes) & store(full) (bytes) \\ \hline \rule{0pt}{1.0\normalbaselineskip}
    $n=4$&5 &         5                              & 2080 & $2.50 \cdot 10^{4}$\\
    &10 &         4                  & 3440 &  $ 7.81 \cdot 10^{7}$\\
    &15 &           4                   & 5840 & $ 2.44 \cdot 10^{11}$\\
    &20 &             4                       & 5800& $7.63 \cdot 10^{14}$  \\
    &25 &              4             & 10920 & $2.38 \cdot 10^{18}$ \\
    \hline \rule{0pt}{1.0\normalbaselineskip}
      $n=6$&5 &        4                              & 2688 & $1.34 \cdot 10^{5}$\\
    &10 &         6                   & 9912  &  $2.26 \cdot 10^{9}$\\
    &15 &           6                   & 13720 & $3.80 \cdot 10^{13}$\\
    &20 &             5                       & 11536& $6.38 \cdot 10^{17}$  \\
    &25 &               4             & 12600  & $1.07 \cdot 10^{22}$ \\
    \end{tabular}
\caption{Completion results on $\Pcal$ for the basket option pricing problem in the Black and Scholes model. Case of uncorrelated assets. \label{compl_results}}
\end{table}

\begin{figure}[ht]
\centering
\includegraphics[width=0.81\textwidth]{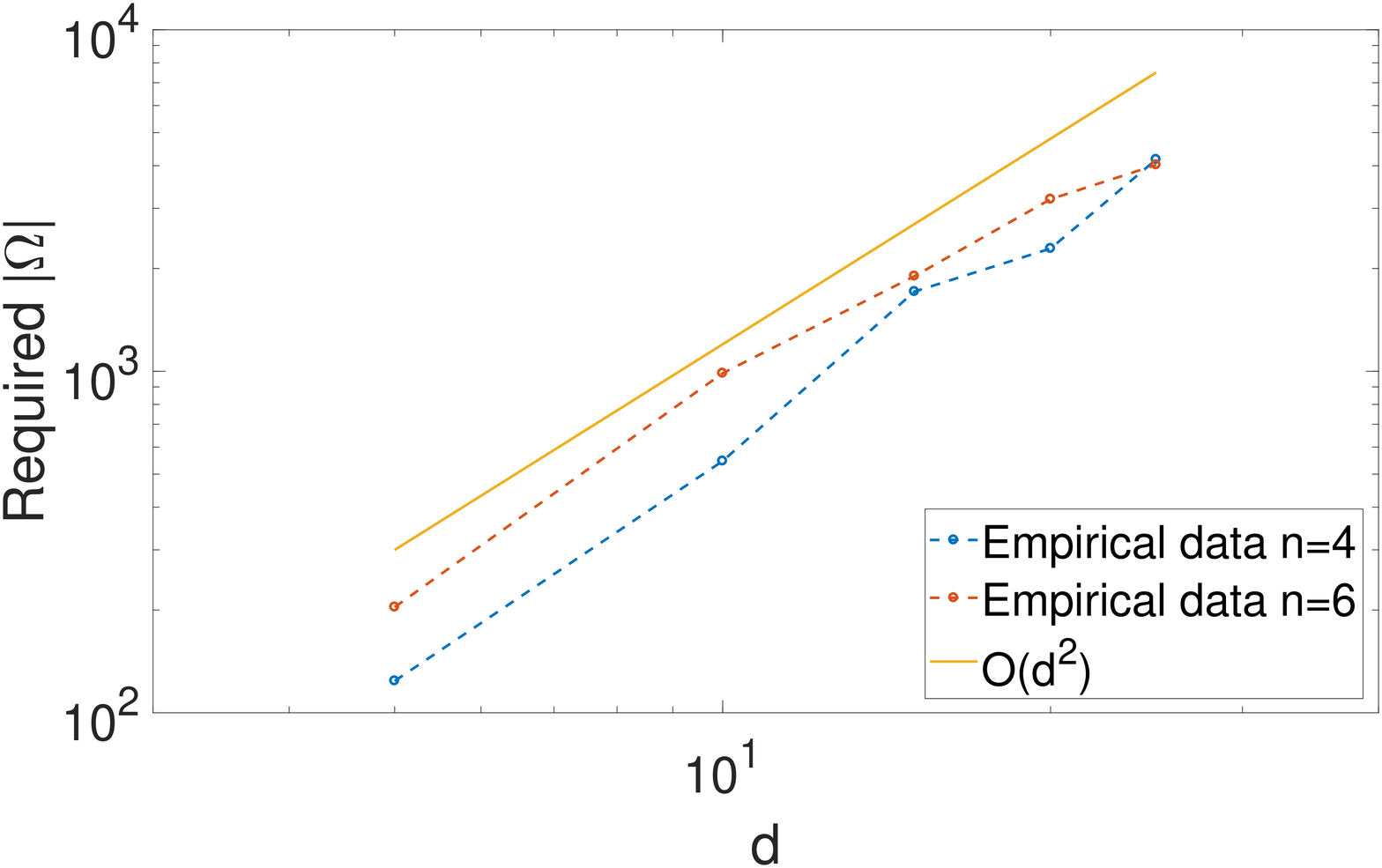}
\caption{Required size of $\Omega$ for the completion to go below $tol = 10^{-2}$ for $n=4$ and $tol = 10^{-3}$ for $n=6$. Case of uncorrelated assets. \label{fig_sizeOmega}}
\end{figure}

It is interesting to analyze the size of the finally obtained set $\Omega$ in Algorithm \ref{adaptive_sampling} for different values of $d$ and $n$ (different sizes of $\Pcal$). Figure \ref{fig_sizeOmega} shows a plot of $|\Omega|$ (final) against $d$ for the two chosen interpolation orders. The graphical representation clearly suggests that the number of sampled entries, i.e. $|\Omega|$, required for the chosen tolerance $tol=10^{-2}$ for a fixed interpolation order $n=4$ and $tol=10^{-3}$ for a fixed $n=6$ is roughly of $O(d^2)$, whereas  the size of the full tensor is $n^d$. On the practical side, this means that by the completion algorithm we can reduce the complexity of the first step of the offline phase from an exponential growth down to a quadratic growth in the dimensionality. The exponential growth typically is referred to as \textit{curse of dimensionality}. The reduction in absolute numbers is already tremendous for $d=5$ and $n=4$, where we observe $|\Omega|=124$ and the full tensor size equals $(n+1)^d=3125$. The compression is dramatic for $n=6$ and $d=25$, namely the numbers of required entries shrinks by a factor of more than $3\times10^{17}$.

As in the previous numerical example, the computation time to build the tensor $\mathcal{C}$ of interpolation coefficients is negligible in the offline phase. Indeed, for all choices of $d$ and $n$ it is less than $0.01$ seconds, for instance $0.0045$ seconds for $n=4$ and $d=5$, and $0.0095$ seconds for $n=6$ and $d=25$.

We now perform the online phase of Algorithm \ref{methodology} in order to see how efficient becomes pricing basket options in the new setting. We start by computing 100 basket option prices via Chebyshev interpolation (combined methodology), choosing random initial asset prices $\mathbf{S_0}$ in the reference hypercube $[1; 1.5]^d$. We then compare the obtained prices with reference prices computed by applying the reference method (Monte Carlo with control variates) with $10^4$ \textsl{new} simulations for $n=4$ and $10^5$ \textsl{new} simulations for $n=6$. In particular, we measure again the maximal absolute error over all computed prices 
\begin{equation*}
\max(|P_{\text{Int}}-P_{\text{Ref}}|),
\end{equation*}
where $P_{\text{Int}}$ is a vector containing all 100 interpolated prices for the different choices of $\mathbf{S_0}$; analogously is $P_{\text{Ref}}$ for the reference method.
The errors together with the computational times are shown in Table \ref{pricesBasket1}. Note that we report again the computational time to compute one single option price.
\begin{table}[ht]
\centering
    \begin{tabular}{ccccc} 
    &$d$  & time reference method (s) & time interpolation (s)& max abs error\\  \hline \rule{0pt}{1.0\normalbaselineskip}
  $n=4$&5&$0.18$ & $0.45 \cdot 10^{-3}$ &   $3.75 \cdot 10^{-3}$ \\ 
  &10 & $0.19$ & $0.64  \cdot 10^{-3}$ &  $5.21 \cdot 10^{-4}$  \\ 
  &15 & $0.20$ & $0.73 \cdot 10^{-3}$ & $4.38 \cdot 10^{-4}$  \\ 
  &20 & $0.20 $ & $1.09 \cdot 10^{-3}$ &  $3.16 \cdot 10^{-4}$ \\ 
  &25 & $0.21 $& $0.97 \cdot 10^{-3}$ &   $2.08 \cdot 10^{-4}$ \\
  \hline  \rule{0pt}{1.0\normalbaselineskip}
    $n=6$&5&1.84& $0.40 \cdot 10^{-3}$&   $5.20 \cdot 10^{-4}$ \\ 
  &10 & 1.91 & $0.61 \cdot 10^{-3}$&  $1.42 \cdot 10^{-4}$  \\ 
  &15 & 1.99 & $0.78 \cdot 10^{-3}$&  $1.02 \cdot 10^{-4}$  \\ 
  &20 & 2.04 & $0.93 \cdot 10^{-3}$&   $1.01  \cdot 10^{-4}$ \\ 
  &25 & 2.10 & $1.04 \cdot 10^{-3}$&   $9.36 \cdot 10^{-5}$ \\
    \end{tabular}
\caption{Basket option prices computed via Chebyshev interpolation (combined methodology) versus MC reference method with $10^4$ simulations for $n=4$ and $10^5$ simulations for $n=6$. Case of uncorrelated assets.
\label{pricesBasket1}}
\end{table}

One can see that the online phase of the new procedure compared to the MC reference method accelerates the computation of a factor between 200 and 400 for $n=4$ and of a factor between 2000 and 4000 for $n=6$. Note that the difference in the acceleration between the two chosen interpolation orders is given by the different numbers of simulations chosen in the MC reference method ($10^4$ for $n=4$ and $10^5$ for $n=6$). Therefore, for both interpolation orders and for all choices of $d$, the acceleration is dramatic. In order to judge the accuracy of our method we have computed the $95 \%$ confidence interval of the reference method, which results to be of a size between $10^{-4}$ and $5 \cdot 10^{-4}$ for all choices of $\mathbf{S}_0$ and $d$ or $n$. This, together with the last column of Table \ref{pricesBasket1}, leads us to the conclusion that the new method is as accurate as the reference MC algorithm. 

\begin{figure}[ht]
\centering
\includegraphics[width=0.86\textwidth]{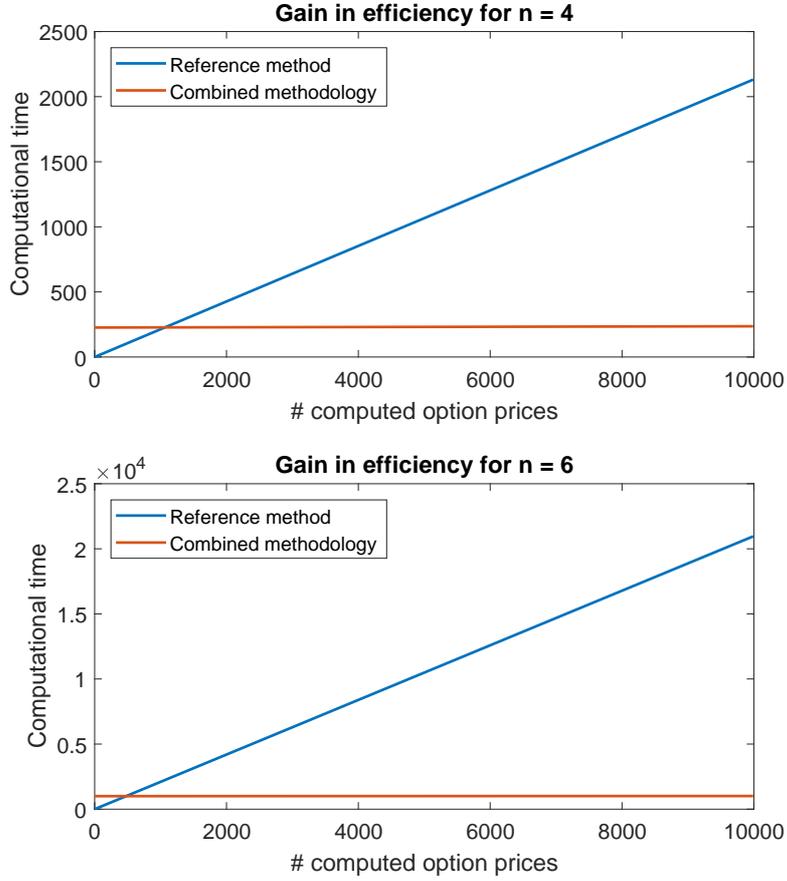}
\caption{Computational time for computing basket option prices. Comparison MC versus combined methodology for $n=4$ and $n=6$. Case of uncorrelated assets. \label{fig_GainEffUncorr}}
\end{figure}

Finally, in Figure \ref{fig_GainEffUncorr} we show the gain in efficiency of the new method when computing basket option prices for $d=25$ and both choices of interpolation orders. In particular, on the x-axis we consider a possible number of computed prices and on the y-axis we present
\begin{enumerate}
\item the computational time of the reference MC method,
\item the computational time of the new combined methodology (offline phase + online phase ), 
\end{enumerate}
required to compute the corresponding amount of prices.

The plots in Figure~\ref{fig_GainEffUncorr} show that after an initial investment the computational time grows very slowly in the number of computed prices for the new method. This is due to the fact that the online phase in Algorithm 4 is very cheap, as shown in the numerical experiments. This proves that the method is useful whenever one can split the task in a pre-computational phase during idle times and a run-time phase where execution is required to be fast. Moreover, it will outperform the reference methods if a large number of prices needs to be computed. The first plot in Figure~\ref{fig_GainEffUncorr} indicates that for the case $n = 4$ it is convenient to use the reference MC method if we want to compute up to $1000$ option prices.  For the case $n = 6 $ the break-even point is already reached with $500$ prices.
\subsubsection{Basket options of correlated assets}
\begin{figure}[ht]
\centering
\includegraphics[width=0.81\textwidth]{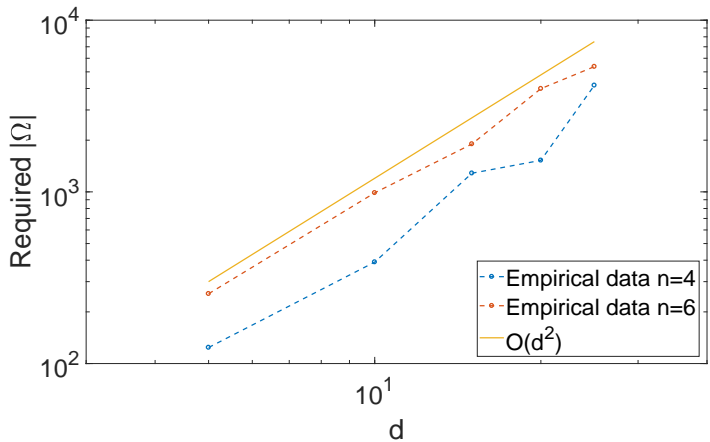}
\caption{Required size of $\Omega$ for the completion to go below $tol = 10^{-2}$ for $n=4$ and $tol = 10^{-3}$ for $n=6$. Case of correlated assets.. \label{fig_sizeOmegaCorr}}
\end{figure}

In this second numerical experiment we repeat the test of the previous subsection but, this time, we consider correlated assets. In particular, we choose again the interpolation orders $n=4$, $n=6$ and the other parameters are given by
\begin{align*}
T=0.25, \quad K=1,\quad r=0, \quad \sigma_i = 0.2 \enskip \forall i, \quad \Sigma= R_d, \quad \omega_i = \frac{1}{d} \enskip \forall i,
\end{align*}
where $R_d$ denotes a random correlation matrix. The free parameters $S_0^i$, $i=1,\cdots,d$ are again contained in [1;1.5].
We perform the offline phase by considering again the set of completion parameters listed in Table \ref{compl_param}. The obtained results of the completion are now in Table \ref{compl_resultsCorr} and Figure \ref{fig_sizeOmegaCorr} shows the required size of $\Omega$ to go below the tolerance $tol = 10^{-2}$ for $n=4$ and $tol = 10^{-3}$ for $n=6$. We notice that the completion results are similar to the case of uncorrelated assets and that $|\Omega|$ scales again like $O(d^2)$. The computational time to construct $ \mathcal{C}$ was again measured to be less than $0.01$ seconds for all choices of $d$ and $n$.

\begin{table}[ht]
\centering
    \begin{tabular}{ccccc}
    &$d$ & final $|\Omega|$& rel err on last $\Omega_C^{new}$ & completion time (s) \\ \hline \rule{0pt}{1.0\normalbaselineskip}
 $n=4$&5  & 124   & $1.86 \cdot 10^{-3}$  & 8.95   \\
          &10 & 390   &$2.19 \cdot 10^{-4}$   & 65.73  \\
          &15 & 1284 & $1.72 \cdot 10^{-7}$  &  118.73  \\
          &20 & 1526 & $2.49 \cdot 10^{-8}$  &  168.20  \\
          &25 & 4172 & $7.52 \cdot 10^{-9}$  & 215.44 \\
    \hline \rule{0pt}{1.0\normalbaselineskip}
  $n=6$&5   & 255&$ 4.40 \cdot 10^{-4}$& 66.54    \\
           &10 & 987 &$2.06 \cdot 10^{-4}$& 200.15  \\
           &15 & 1900& $1.79 \cdot 10^{-7}$&  432.58  \\
           &20 & 3990& $1.82 \cdot 10^{-8}$&  852.13  \\
           &25 & 5364 & $2.88 \cdot 10^{-7}$& 1335.76 \\
    \end{tabular}
    
    \vspace{0.5 cm}
    
        \begin{tabular}{ccccc}
    &$d$ &  max $r_{\mu}$ reached & store(TT) (bytes)& store(full) (bytes) \\ \hline \rule{0pt}{1.0\normalbaselineskip}
  $n=4$&5   &    5     & 2320 &$ 2.50 \cdot 10^{4}$\\
           &10 &    3      &  2440 &  $ 7.81 \cdot 10^{7}$\\
           &15 &     5     & 8960 & $ 2.44 \cdot 10^{11}$\\
           &20 &     4    & 7040& $ 7.63 \cdot 10^{14}$  \\
           &25 &     3    & 8040& $  2.38 \cdot 10^{18}$\\
    \hline \rule{0pt}{1.0\normalbaselineskip}
    $n=6$&5 & 5     & 2520 &$1.34 \cdot 10^{5}$\\
    &10 &  5     &  6832 &  $2.26 \cdot  10^{9}$\\
    &15 &    4      & 6664 & $ 3.80  \cdot  10^{13}$\\
    &20 &      4    & 12040& $ 6.38 \cdot 10^{17}$  \\
    &25 &      4    & 8960& $1.07 \cdot 10^{22}$\\
    \end{tabular}
\caption{Completion results on $\Pcal$ for the basket option pricing problem in the Black and Scholes model. Case of correlated assets.   \label{compl_resultsCorr}}
\end{table}

The online phase is performed similarly to the previous chapter, in particular we compute again 100 prices using the new method and the reference one. The MC parameters are set as before and the results are shown in Table \ref{pricesBasket1Corr}.
\begin{table}[ht]
\centering
    \begin{tabular}{ccccc} 
   & $d$  & time reference method (s)& time interpolation (s)& max abs error\\  \hline \rule{0pt}{1.0\normalbaselineskip}
   $n=4$&5&0.18 & $0.50 \cdot 10^{-3}$ & $1.39 \cdot 10^{-3}$ \\ 
            &10 & 0.20 & $0.56  \cdot 10^{-3}$ & $4.82 \cdot 10^{-4}$ \\ 
            &15 & 0.20 & $0.70 \cdot 10^{-3}$ & $2.82 \cdot 10^{-4}$ \\ 
            &20 & 0.23  & $0.91  \cdot 10^{-3}$ & $2.93 \cdot 10^{-4}$ \\ 
            &25 & 0.23  & $ 1 \cdot 10^{-3}$ & $4.30 \cdot 10^{-4}$ \\ 
  \hline \rule{0pt}{1.0\normalbaselineskip}
   $n=6$&5&1.85 & $0.38 \cdot 10^{-3}$ & $3.55 \cdot 10^{-4}$ \\ 
 	    &10 & 1.90 & $0.57 \cdot 10^{-3}$ & $5.58 \cdot 10^{-4}$ \\ 
  	    &15 & 1.99 & $0.74 \cdot 10^{-3}$ & $1.39 \cdot 10^{-4}$ \\ 
  	   &20 & 2.06  & $0.90 \cdot 10^{-3}$ & $1.41  \cdot 10^{-4}$ \\ 
  	   &25 & 2.15  & $0.96 \cdot 10^{-3}$ & $9.28 \cdot 10^{-5}$ \\ 
    \end{tabular}
    \caption{Basket option prices computed via Chebyshev interpolation (combined methodology) versus MC reference method with $10^4$ simulations for $n=4$ and $10^5$ simulations for $n=6$. Case of correlated assets.
\label{pricesBasket1Corr}}
\end{table}
The performance of the new method in terms of accuracy and computational efficiency is similar to the one observed in the case of uncorrelated assets. To summarize, the new methodology achieves a very good performance for uncorrelated as well as for correlated assets.

\subsubsection{Rank structure of $\Pcal$}\label{analysisP}
In this section we qualitatively analyze the rank structure of the tensor $\Pcal$.
For simplicity, we perform this analysis for the standard Monte Carlo approach (without any variance reduction technique). Assume that we have already simulated the realizations of the correlated geometric Brownian motions stored in the matrix $M$ (Algorithm \ref{SimCorrGBM}). Then, the price in the point $\mathbf{S}_0$ is given by the function
\begin{align*}
&p : D \to \mathbb{R}, \\
&p(\mathbf{S_0}) \defby \frac{e^{-rT}}{N_S} \sum_{n=1}^{N_S} \big [ w^T (\mathbf{S}_0 \circ M(n, :)^T)-K  \big ]^+,
\end{align*}
where $D$ is the hyper-rectangular domain for the interpolation, $M(n,:)$ is the $n$-th row of $M$ and $N_S$ is the number of Monte Carlo simulations. This expression can be rewritten in the form 
\begin{equation*}
p(\mathbf{S_0}) = \frac{e^{-rT}}{N_S}\sum_{n=1}^{N_S} \Big ( \sum_{i=1}^d \alpha_i(n) S_0^i  -K \Big )^+,
\end{equation*}
where the $\alpha_i (n)$'s are coefficients multiplying $S_0^i$ depending on the $n$-th simulation and on the $i$-th weight $\omega_i$.
The function $p$ is piecewise affine in the variables $S_0^i$.

To explore the rank structure of $\Pcal$ let us consider the case of a single Monte Carlo simulation $N_S=1$. Then $p$ is of the form
\begin{equation*}
p(\mathbf{S_0}) = e^{-rT} \Big ( \sum_{i=1}^d \alpha_i S_0^i  -K \Big )^+.
\end{equation*}
Now we analyze three different cases. First, consider the case where the price is positive for any $\mathbf{S}_0$ in the hyper-rectangular $D$. Here, $p$ is affine. This implies that the TT ranks are bounded by $d$. This follows from the fact that the CP rank (rank of the Canonical Polyadic Decomposition, see~\cite{kolda2009tensor}) of $\Pcal$,  which is an upper bound for each $r_{\mu}$ in the TT ranks (see~\cite{hackbusch2012tensor}), is equal to $d$.
Second, if we observe a vanishing price for all $\mathbf{S}_0$ in the hyper-rectangular, then $\Pcal$ is the zero-tensor, which has rank $0$.
These two cases obviously yield a low-rank structure of $\Pcal$, a favorable case for the new combined methodology.

In the third case where $p$ is only piecewise affine the situation is more complex and to gain an intuition we consider the case $d=2$, where $p$ is of the form 
\begin{equation*}
p(S_0^1, S_0^2) = e^{-rT} (\alpha_1 S_0^1+\alpha_2 S_0^2 -K)^+,
\end{equation*}
on a squared domain $D$.
Now, define the set
\begin{equation*}
L \defby \{(S_0^1, S_0^2) \in D \enskip | \enskip \alpha_1 S_0^1+\alpha_2 S_0^2 -K = 0 \}.
\end{equation*}
When $L$ intersects the domain $D$ it cuts it in two regions. Only if $\alpha_1, \alpha_2$ and $K$ are of a specific form that leads L to be the diagonal of $D$, the rank of $\Pcal$ is almost full. In the Monte Carlo simulation context, this special case is very unlikely. In all other cases, $\Pcal$ exhibits a lower rank structure. In particular, we expect the rank to be the lower the more the sizes of the two regions differ.

\begin{figure}[ht]
\centering
\includegraphics[width=0.32\textwidth]{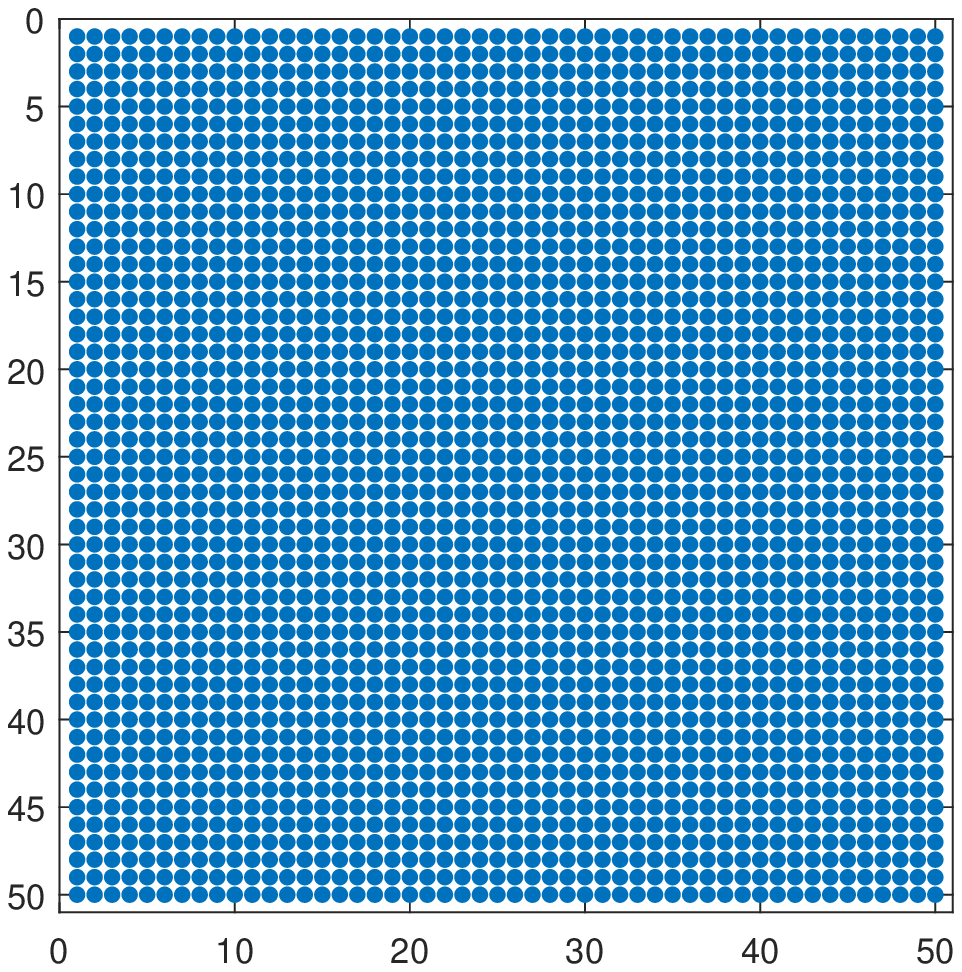}
\includegraphics[width=0.32\textwidth]{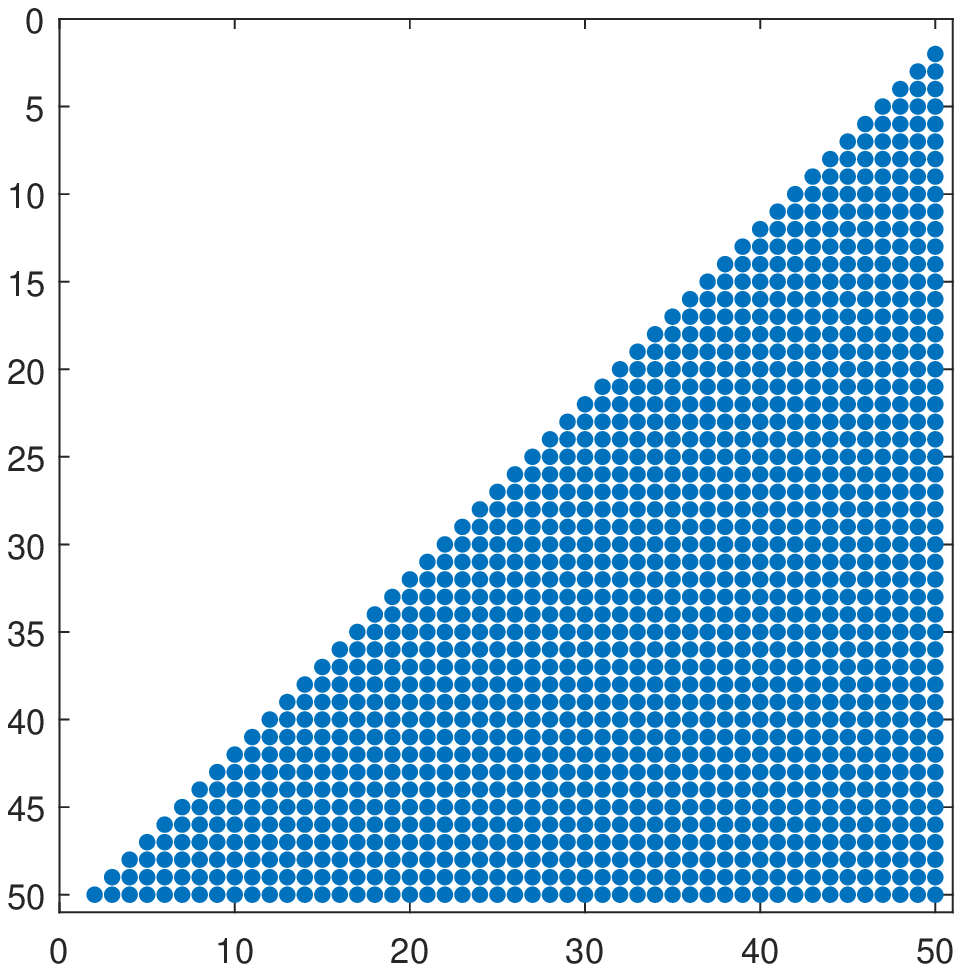}
\includegraphics[width=0.32\textwidth]{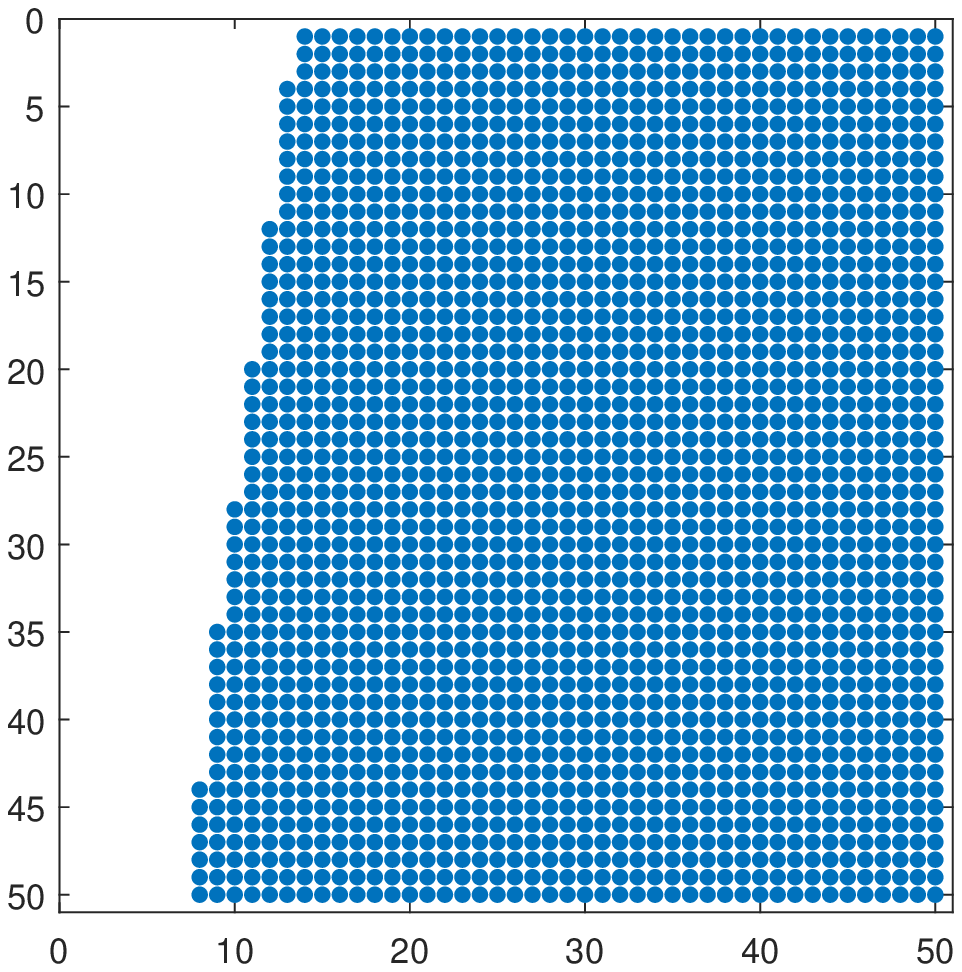}
\caption{Sparsity patterns and ranks for evaluated $p$ on $D=[1;1.5]^2$ for different values of $(\alpha_1, \alpha_2)$. \textsl{Left:} 
$(\alpha_1, \alpha_2)=(0.9,0.8)$ and $\text{rank}=2$. \textsl{Center:} $(\alpha_1, \alpha_2)=(0.4,0.4)$ and $\text{rank}=49$. 
 \textsl{Right:} $(\alpha_1, \alpha_2)=(0.1,0.8)$ and $\text{rank}=8$. \label{sparsities}}
\end{figure}

In order to visualize these findings we consider three different pairs $(\alpha_1, \alpha_2)$ together with $r=0$, $K=1$ and evaluate the corresponding $p$ on the discretized $D=[1;1.5]^2$ using 50 equidistant points in each direction. Figure \ref{sparsities} shows the sparsity pattern and the rank of the obtained matrices $\Pcal$.

\begin{figure}[ht]
\centering
\includegraphics[width=0.49\textwidth]{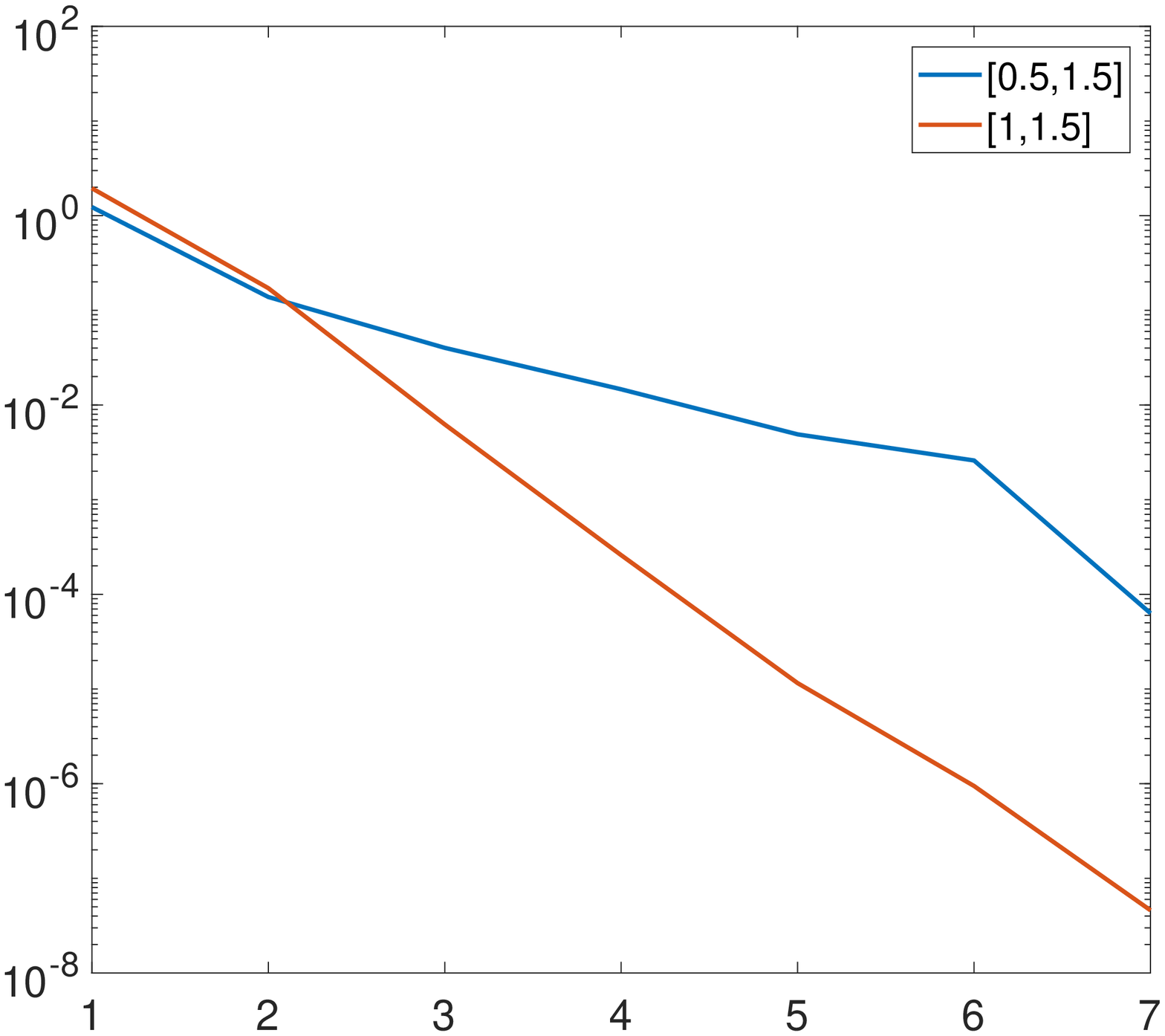}
\includegraphics[width=0.49\textwidth]{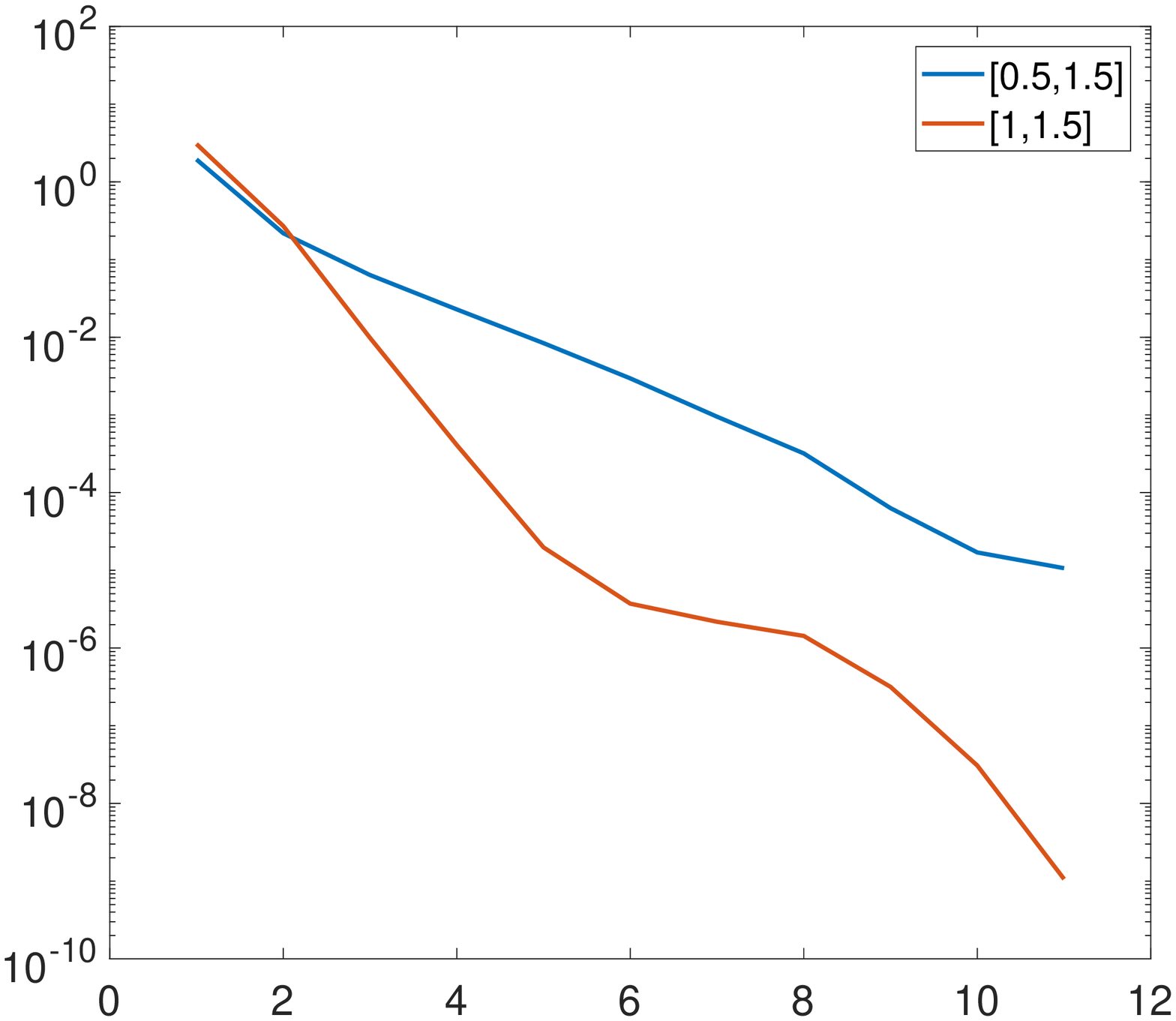}
\includegraphics[width=0.49\textwidth]{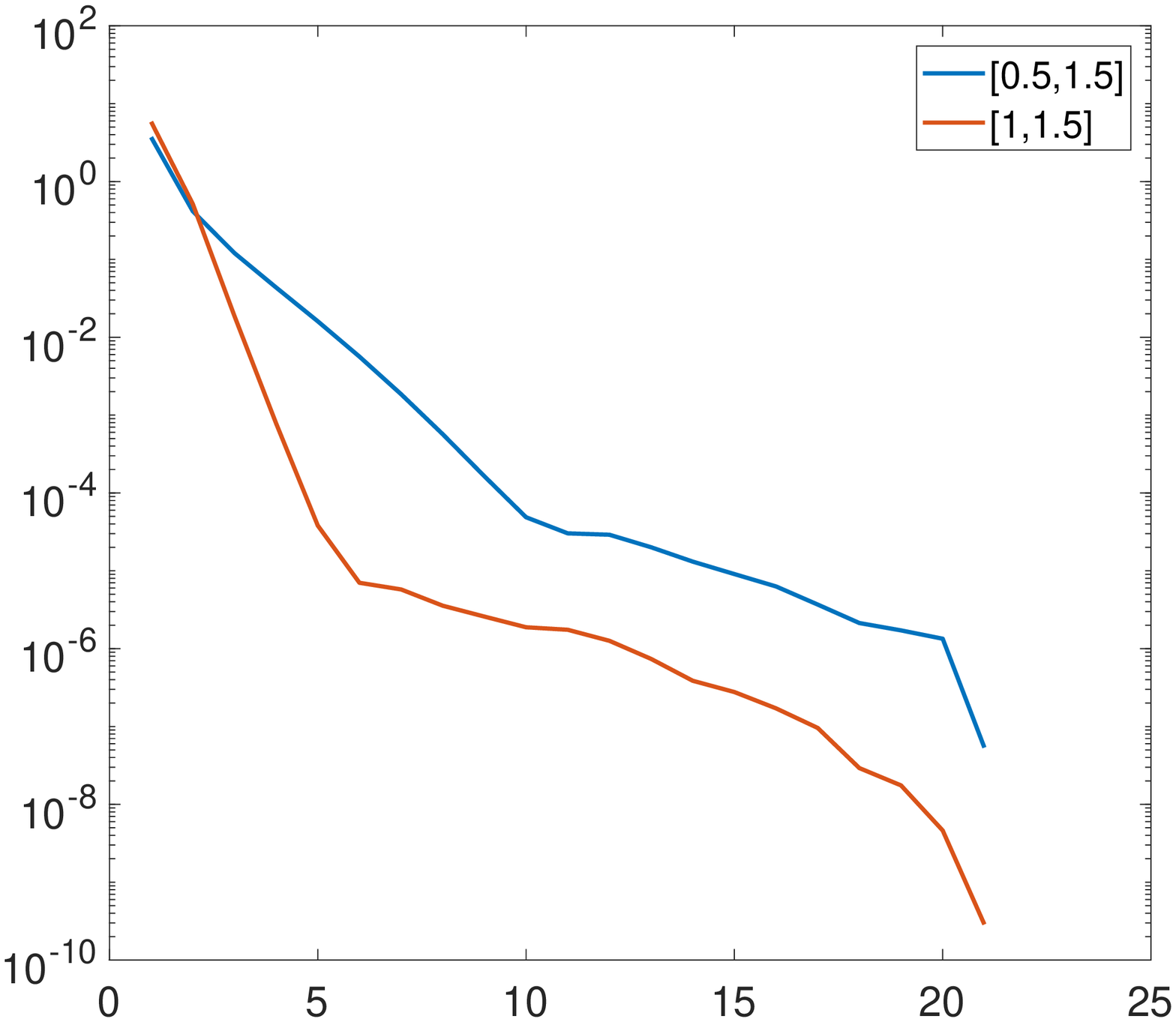}
\caption{Singular value decay of the matrix $\Pcal$ for sampling intervals $[0,5;1,5]$ and $[1;1.5]$ for different interpolation orders: \textsl{Top left:} $n=6$, \textsl{Top right:} $n=10$, \textsl{Bottom:} $n=20$.  \label{SVDdecay}}
\end{figure}

This qualitative explanation indicates that the rank structure of $\Pcal$ depends on $D$. We expect the rank to be lower for domains $D$ with an asymmetry with respect to the strike $K$. 
Next we construct $\Pcal$ as in the experiments of Section~\ref{numerical exp BS} for $K=1$,  $d=2$ and different interpolation orders $n$ for both $D=[0.5;1.5]^2$ and $D=[1;1.5]^2$. In particular, we first construct the matrix $M$ via Algorithm \ref{SimCorrGBM} with $10^5$ simulations and subsequently compute $\Pcal$ using Algorithm \ref{BasketNdSim}. In Figure \ref{SVDdecay} we display the decay of the singular values for all treated cases. As expected, the decay is faster for $D=[1;1.5]^2$. However, also for $D=[0.5;1.5]^2$ the decay of the singular values is reasonably fast. This implies that the new methodology would be still beneficial in this case. 

\section{Summary and future work}
\label{sec-concl}
We have presented a unified approach to efficiently compute parametric option prices. The starting point of our methodology was the Chebyshev interpolation technique developed in \cite{gass2016chebyshev}, which we briefly summarized in Section \ref{chebyshevPOP}. We refined both the offline and the online phase to treat high-dimensional problems with parameter spaces up to dimension $25$. We have exploited the low-rank structure of the tensors involved in the interpolation procedure, which have been stored in TT format (summarized in Section \ref{sec-ttformat}). In particular, we have developed a completion technique (explained in Section~\ref{sec-completion}) which allows us the construct the tensor $\Pcal$, containing the option prices in the Chebshev tensor grid. All ingredients have been efficiently assembled to finally build a combined methodology, explained in Section \ref{sec-combined methodology}. 

In the second part of the paper, Section \ref{section3}, we have tested our approach in two different concrete option pricing settings: We have treated the American option pricing problem in the Heston model (Section \ref{sectionHeston}) and the European basket option pricing problem in the $d$-dimensional Black and Scholes model (Section \ref{numerical exp BS}). Both examples show that our approach allows for a substantial gain in efficiency, while maintaining very accurate results, whose precision is comparable to the one of the considered reference methods. For instance, the interpolation of American option prices in $5$ parameters accelerates the procedure by a factor of $75$, when compared to the FD reference method \cite{haentjens2015adi}. For basket option pricing with $25$ underlyings the efficiency gain reaches factors up to $4000$. See Tables \ref{pricesAmerican}, \ref{pricesBasket1}, \ref{pricesBasket1Corr} and Figure \ref{fig_GainEffUncorr} for further results. Finally, for both examples we qualitatively investigated the rank structure of $\Pcal$, which confirmed that our initial low-rank assumption was indeed reasonable. 
For instance, for the American put, we obtain a compression factor of $115$ of the completed tensor $\Pcal$ with respect to the full one, with a relative error in the $5$th digit only, see Table \ref{completionAmerican}. For the basket option the full tensor containing prices in the Chebyshev grid is too large to be computed, however in Section \ref{analysisP} it is qualitatively explained why $\Pcal$ is expected to have a low-rank structure. This is also confirmed by the compression rates observed in Tables \ref{compl_results} and \ref{compl_resultsCorr} that go up to $3 \times 10^{17}$.

Seen the promising performance of this new approach and considering the fact that this methodology can be easily tailored to different problem settings, we expect it to be applicable in several domains in finance. 
For instance, 
pricing, calibration and sensitivity analysis in equity markets, fixed income and credit, and
parameter uncertainty quantification are some of the possible domains of application.

\bibliographystyle{siamplain}
\bibliography{references}

\begin{thebibliography}{10}

\bibitem{Bachmayr2017}
{\sc M.~Bachmayr and A.~Cohen}, {\em Kolmogorov widths and low-rank
  approximations of parametric elliptic {PDE}s}, Math. Comp., 86 (2017),
  pp.~701--724, \url{http://dx.doi.org/10.1090/mcom/3132}.

\bibitem{brett2006algorithm}
{\sc B.~W. Bader and T.~G. Kolda}, {\em Algorithm 862: {MATLAB} tensor classes
  for fast algorithm prototyping}, ACM Transactions on Mathematical Software,
  32 (2006), pp.~635--653, \url{http://dx.doi.org/10.1145/1186785.1186794}.

\bibitem{TTB_Software}
{\sc B.~W. Bader, T.~G. Kolda, et~al.}, {\em Matlab tensor toolbox version
  2.6}.
\newblock Available online, February 2015,
  \url{http://www.sandia.gov/~tgkolda/TensorToolbox/}.

\bibitem{Ballani2015}
{\sc J.~Ballani and L.~Grasedyck}, {\em Hierarchical tensor approximation of
  output quantities of parameter-dependent {PDE}s}, SIAM/ASA J. Uncertain.
  Quantif., 3 (2015), pp.~852--872, \url{http://dx.doi.org/10.1137/140960980}.

\bibitem{BarreraCrepeyDialloFortGobetStazhynski2019}
{\sc D.~Barrera, S.~Cr\'epey, B.~Diallo, G.~Fort, E.~Gobet, and U.~Stazhynski},
  {\em {Stochastic approximation schemes for economic capital and risk margin
  computations}}.
\newblock Forthcoming in ESAIM: Proceedings and Surveys,
  https://math.maths.univ-evry.fr/crepey/papers/SA-EC-RM.pdf, 2019.

\bibitem{BayerSiebenmorgenTempone2017}
{\sc C.~Bayer, M.~Siebenmorgen, and R.~Tempone}, {\em Smoothing the payoff for
  efficient computation of basket option prices}, Quant. Finance, 18 (2018),
  pp.~491--505, \url{http://dx.doi.org/10.1080/14697688.2017.1308003}.

\bibitem{BurkovskaGlauMahlstedtWohlmuth2017}
{\sc O.~Burkovska, K.~Glau, M.~Mahlstedt, and B.~Wohlmuth}, {\em {Complexity
  reduction for calibration of American options}}.
\newblock Forthcoming in J. Comput. Finance, https://arxiv.org/abs/1611.06452,
  2017.

\bibitem{BurkovskaHaasdonkSalomonWohlmuth2015}
{\sc O.~Burkovska, B.~Haasdonk, J.~Salomon, and B.~Wohlmuth}, {\em Reduced
  basis methods for pricing options with the {B}lack-{S}choles and {H}eston
  models}, SIAM J. Financial Math., 6 (2015), pp.~685--712,
  \url{http://dx.doi.org/10.1137/140981216}.

\bibitem{CapriottiJiangMacrina2017}
{\sc L.~Capriotti, Y.~Jiang, and A.~Macrina}, {\em A{AD} and least-square
  {M}onte {C}arlo: fast {B}ermudan-style options and {XVA} {G}reeks},
  Algorithmic Finance, 6 (2017), pp.~35--49,
  \url{http://dx.doi.org/10.3233/af-170201}.

\bibitem{ContLantosPironneau2011}
{\sc R.~Cont, N.~Lantos, and O.~Pironneau}, {\em A reduced basis for option
  pricing}, SIAM J. Financial Math., 2 (2011), pp.~287--316,
  \url{http://dx.doi.org/10.1137/10079851X}.

\bibitem{Dahmen2016}
{\sc W.~Dahmen, R.~DeVore, L.~Grasedyck, and E.~S\"{u}li}, {\em Tensor-sparsity
  of solutions to high-dimensional elliptic partial differential equations},
  Found. Comput. Math., 16 (2016), pp.~813--874,
  \url{http://dx.doi.org/10.1007/s10208-015-9265-9}.

\bibitem{dempster2018high}
{\sc M.~A.~H. Dempster, J.~Kanniainen, J.~Keane, and E.~Vynckier}, {\em
  High-Performance Computing in Finance: Problems, Methods, and Solutions},
  Chapman \& Hall/CRC, 1st~ed., 2018.

\bibitem{duffy2006finite}
{\sc D.~J. Duffy}, {\em {Finite Difference Methods in Financial Engineering: A
  Partial Differential Equation Approach}}, Wiley Finance Series, John Wiley \&
  Sons, Ltd., Chichester, 2006, \url{http://dx.doi.org/10.1002/9781118673447}.

\bibitem{gass2016chebyshev}
{\sc M.~Ga\ss, K.~Glau, M.~Mahlstedt, and M.~Mair}, {\em Chebyshev
  interpolation for parametric option pricing}, Finance Stoch., 22 (2018),
  pp.~701--731, \url{http://dx.doi.org/10.1007/s00780-018-0361-y}.

\bibitem{Giles2015}
{\sc M.~B. Giles}, {\em Multilevel {M}onte {C}arlo methods}, Acta Numer., 24
  (2015), pp.~259--328, \url{http://dx.doi.org/10.1017/S096249291500001X}.

\bibitem{GilesXia2017}
{\sc M.~B. Giles and Y.~Xia}, {\em Multilevel {M}onte {C}arlo for exponential
  {L}\'{e}vy models}, Finance Stoch., 21 (2017), pp.~995--1026,
  \url{http://dx.doi.org/10.1007/s00780-017-0341-7}.

\bibitem{glasserman2004monte}
{\sc P.~Glasserman}, {\em Monte {C}arlo methods in financial engineering},
  vol.~53 of Applications of Mathematics, Springer-Verlag, New York, 2004.
\newblock Stochastic Modelling and Applied Probability.

\bibitem{grasedyck2013}
{\sc L.~Grasedyck, D.~Kressner, and C.~Tobler}, {\em A literature survey of
  low-rank tensor approximation techniques}, GAMM-Mitt., 36 (2013), pp.~53--78,
  \url{http://dx.doi.org/10.1002/gamm.201310004}.

\bibitem{griebel2010dimension}
{\sc M.~Griebel and M.~Holtz}, {\em Dimension-wise integration of
  high-dimensional functions with applications to finance}, J. Complexity, 26
  (2010), pp.~455--489, \url{http://dx.doi.org/10.1016/j.jco.2010.06.001}.

\bibitem{hackbusch2012tensor}
{\sc W.~Hackbusch}, {\em Tensor spaces and numerical tensor calculus}, vol.~42
  of Springer Series in Computational Mathematics, Springer, Heidelberg, 2012,
  \url{http://dx.doi.org/10.1007/978-3-642-28027-6}.

\bibitem{haentjens2015adi}
{\sc T.~Haentjens and K.~J. in't Hout}, {\em A{DI} schemes for pricing
  {A}merican options under the {H}eston model}, Appl. Math. Finance, 22 (2015),
  pp.~207--237, \url{http://dx.doi.org/10.1080/1350486X.2015.1009129}.

\bibitem{hashemi2016chebfun}
{\sc B.~Hashemi and L.~N. Trefethen}, {\em Chebfun in three dimensions}, SIAM
  J. Sci. Comput., 39 (2017), pp.~C341--C363,
  \url{http://dx.doi.org/10.1137/16M1083803}.

\bibitem{Hesthaven2015}
{\sc J.~S. Hesthaven, G.~Rozza, and B.~Stamm}, {\em Certified reduced basis
  methods for parametrized partial differential equations}, SpringerBriefs in
  Mathematics, Springer, Cham; BCAM Basque Center for Applied Mathematics,
  Bilbao, 2016, \url{http://dx.doi.org/10.1007/978-3-319-22470-1}.
\newblock BCAM SpringerBriefs.

\bibitem{heston1993closed}
{\sc S.~L. Heston}, {\em A closed-form solution for options with stochastic
  volatility with applications to bond and currency options}, Review of
  Financial Studies, 6 (1993), pp.~327--343.

\bibitem{HilberReichSchwabWinter2009}
{\sc N.~Hilber, N.~Reich, C.~Schwab, and C.~Winter}, {\em Numerical methods for
  {L}\'{e}vy processes}, Finance Stoch., 13 (2009), pp.~471--500,
  \url{http://dx.doi.org/10.1007/s00780-009-0100-5}.

\bibitem{HilberReichmannSchwabWinter2013}
{\sc N.~Hilber, O.~Reichmann, C.~Schwab, and C.~Winter}, {\em {Computational
  Methods for Quantitative Finance}}, Springer Finance, Springer, Heidelberg,
  2013, \url{http://dx.doi.org/10.1007/978-3-642-35401-4}.
\newblock Finite element methods for derivative pricing.

\bibitem{holtz2011sparse}
{\sc M.~Holtz}, {\em Sparse grid quadrature in high dimensions with
  applications in finance and insurance}, vol.~77 of Lecture Notes in
  Computational Science and Engineering, Springer-Verlag, Berlin, 2011,
  \url{http://dx.doi.org/10.1007/978-3-642-16004-2}.

\bibitem{HoutToivanen2016}
{\sc K.~in't Hout and J.~Toivanen}, {\em Application of operator splitting
  methods in finance}, Sci. Comput., Springer, Cham, 2016, pp.~541--575.

\bibitem{Khoromskij2018}
{\sc B.~N. Khoromskij}, {\em Tensor numerical methods in scientific computing},
  vol.~19 of Radon Series on Computational and Applied Mathematics, De Gruyter,
  Berlin, 2018.

\bibitem{Khoromskij2011c}
{\sc B.~N. Khoromskij and C.~Schwab}, {\em Tensor-structured {G}alerkin
  approximation of parametric and stochastic elliptic {PDE}s}, SIAM J. Sci.
  Comput., 33 (2011), pp.~364--385, \url{http://dx.doi.org/10.1137/100785715}.

\bibitem{kolda2009tensor}
{\sc T.~G. Kolda and B.~W. Bader}, {\em Tensor decompositions and
  applications}, SIAM Rev., 51 (2009), pp.~455--500,
  \url{http://dx.doi.org/10.1137/07070111X}.

\bibitem{kressner2014low}
{\sc D.~Kressner, M.~Steinlechner, and B.~Vandereycken}, {\em Low-rank tensor
  completion by {R}iemannian optimization}, BIT, 54 (2014), pp.~447--468,
  \url{http://dx.doi.org/10.1007/s10543-013-0455-z}.

\bibitem{kressner2016preconditioned}
{\sc D.~Kressner, M.~Steinlechner, and B.~Vandereycken}, {\em Preconditioned
  low-rank {R}iemannian optimization for linear systems with tensor product
  structure}, SIAM J. Sci. Comput., 38 (2016), pp.~A2018--A2044,
  \url{http://dx.doi.org/10.1137/15M1032909}.

\bibitem{kressner2011low-rank}
{\sc D.~Kressner and C.~Tobler}, {\em Low-rank tensor {K}rylov subspace methods
  for parametrized linear systems}, SIAM J. Matrix Anal. Appl., 32 (2011),
  pp.~1288--1316, \url{http://dx.doi.org/10.1137/100799010}.

\bibitem{LEcuyer2009}
{\sc P.~L'Ecuyer}, {\em Quasi-{M}onte {C}arlo methods with applications in
  finance}, Finance Stoch., 13 (2009), pp.~307--349,
  \url{http://dx.doi.org/10.1007/s00780-009-0095-y}.

\bibitem{MatacheNitscheSchwab2005}
{\sc A.-M. Matache, P.-A. Nitsche, and C.~Schwab}, {\em Wavelet {G}alerkin
  pricing of {A}merican options on {L}\'{e}vy driven assets}, Quant. Finance, 5
  (2005), pp.~403--424, \url{http://dx.doi.org/10.1080/14697680500244478}.

\bibitem{MayerhoferUrban2017}
{\sc A.~Mayerhofer and K.~Urban}, {\em A reduced basis method for parabolic
  partial differential equations with parameter functions and application to
  option pricing}, J. Comput. Finance, 20 (2017), pp.~71--106.

\bibitem{orus2014practical}
{\sc R.~Or\'{u}s}, {\em A practical introduction to tensor networks: matrix
  product states and projected entangled pair states}, Ann. Physics, 349
  (2014), pp.~117--158, \url{http://dx.doi.org/10.1016/j.aop.2014.06.013}.

\bibitem{oseledets2011}
{\sc I.~V. Oseledets}, {\em Tensor-train decomposition}, SIAM J. Sci. Comput.,
  33 (2011), pp.~2295--2317, \url{http://dx.doi.org/10.1137/090752286}.

\bibitem{Patera2006}
{\sc A.~T. Patera and G.~Rozza}, {\em Reduced basis approximation and a
  posteriori error estimation for parametrized partial differential equations},
  Tech. Report Version 1.0, MIT 2006--2007, to appear in (tentative rubric) MIT
  Pappalardo Graduate Monographs in Mechanical Engineering, Massachusetts
  Institute of Technology, 2006,
  \url{http://augustine.mit.edu/methodology/bookParts/Patera_Rozza_bookPartI_BV1.pdf}.

\bibitem{QuarteroniManzoniNegri2016}
{\sc A.~Quarteroni, A.~Manzoni, and F.~Negri}, {\em Reduced basis methods for
  partial differential equations}, vol.~92 of Unitext, Springer, Cham, 2016,
  \url{http://dx.doi.org/10.1007/978-3-319-15431-2}.
\newblock An introduction, La Matematica per il 3+2.

\bibitem{ReisingerWissmann2016}
{\sc C.~Reisinger and R.~Wissmann}, {\em { Numerical valuation of derivatives
  in high-dimensional settings via PDE expansions}}, J. Comput. Finance, 18
  (2015), pp.~95--127.

\bibitem{SachsSchu2008}
{\sc E.~W. Sachs and M.~Schu}, {\em Reduced order models ({POD}) for
  calibration problems in finance}, in Numerical mathematics and advanced
  applications, Springer, Berlin, 2008, pp.~735--742.

\bibitem{SchneiderUschmajew2014}
{\sc R.~Schneider and A.~Uschmajew}, {\em Approximation rates for the
  hierarchical tensor format in periodic {S}obolev spaces}, J. Complexity, 30
  (2014), pp.~56--71, \url{http://dx.doi.org/10.1016/j.jco.2013.10.001}.

\bibitem{steinlechner2016riemannian}
{\sc M.~Steinlechner}, {\em Riemannian optimization for high-dimensional tensor
  completion}, SIAM J. Sci. Comput., 38 (2016), pp.~S461--S484,
  \url{http://dx.doi.org/10.1137/15M1010506}.

\bibitem{steinlechner2016thesis}
{\sc M.~Steinlechner}, {\em Riemannian Optimization for Solving
  High-Dimensional Problems with Low-Rank Tensor Structure}, PhD thesis,
  {\'E}cole polytechnique f{\'e}d{\'e}rale de Lausanne, 2016.

\bibitem{uschmajew2015greedy}
{\sc A.~Uschmajew and B.~Vandereycken}, {\em {Greedy rank updates combined with
  Riemannian descent methods for low-rank optimization}}, in Proceedings of the
  2015 International Conference on Sampling Theory and Applications (SampTA),
  IEEE, 2015, pp.~420--424.

\end{thebibliography}

\end{document}